\documentclass[journal,comsoc, 12pt, onecolumn, draftclsnofoot]{IEEEtran}
\usepackage{cite}
\usepackage{amsmath,amssymb,amsfonts,helvet}
\usepackage[cmintegrals]{newtxmath}
\usepackage{bbm}
\usepackage{algorithmic}
\usepackage{graphicx}
\usepackage{textcomp}
\usepackage{soul}
\def\BibTeX{{\rm B\kern-.05em{\sc i\kern-.025em b}\kern-.08em
    T\kern-.1667em\lower.7ex\hbox{E}\kern-.125emX}}
    
\usepackage[
  pass,
]{geometry}
\usepackage{graphicx,epsfig}
\usepackage{subcaption}
\usepackage{caption,setspace}
\usepackage{cite}
\usepackage{url}
\usepackage[abs]{overpic}
\usepackage{footmisc}
\usepackage[alwaysadjust]{paralist}
\usepackage{hyperref}
\usepackage{multirow}
\usepackage{color, colortbl}
\usepackage{soul}
\usepackage{listings}
\usepackage{color}
\usepackage{makecell}

\usepackage{acro}
\DeclareAcronym{3GPP}{
  short=3GPP,
  long=3rd generation partnership project
}
\DeclareAcronym{ADC}{
  short=ADC,
  long=analog-to-digital converter
}
\DeclareAcronym{AMP}{
  short=AMP,
  long=approximate message passing
}
\DeclareAcronym{ANN}{
  short=ANN,
  long=artificial neural network
}
\DeclareAcronym{AoA}{
  short=AoA,
  long=angle-of-arrival
}
\DeclareAcronym{AoD}{
  short=AoD,
  long=angle-of-departure
}
\DeclareAcronym{APS}{
  short=APS,
  long=azimuth power spectrum
}
\DeclareAcronym{AR}{
  short=AR,
  long=augmented reality
}
\DeclareAcronym{AV}{
  short=AV,
  long=autonomous vehicle
}
\DeclareAcronym{BM}{
  short=BM,
  long=beam management
}
\DeclareAcronym{BS}{
  short=BS,
  long=base station
}
\DeclareAcronym{BSM}{
  short=BSM,
  long=basic safety message
}
\DeclareAcronym{CDF}{
  short=CDF,
  long=cumulative distribution function
}
\DeclareAcronym{CP}{
  short=CP,
  long=cyclic-prefix
}
\DeclareAcronym{CSI-RS}{
  short=CSI-RS,
  long=Channel State Information Reference Signal
}
\DeclareAcronym{CSI}{
  short=CSI,
  long=Channel State Information
}
\DeclareAcronym{DFT}{
  short=DFT,
  long=discrete Fourier transform
}
\DeclareAcronym{EKF}{
  short=EKF,
  long=extended Kalman filter
}
\DeclareAcronym{DSRC}{
  short=DSRC,
  long=dedicated short-range communication
}
\DeclareAcronym{FDD}{
  short=FDD,
  long=frequency division duplex
}
\DeclareAcronym{FMCW}{
  short=FMCW,
  long=frequency modulated continuous wave
}
\DeclareAcronym{FoV}{
  short=FoV,
  long=field-of-view
}
\DeclareAcronym{GNSS}{
  short=GNSS,
  long=global navigation satellite system
}
\DeclareAcronym{IA}{
  short=IA,
  long=initial access
}
\DeclareAcronym{IMU}{
  short=IMU,
  long=inertial measurement unit
}
\DeclareAcronym{lidar}{
  short=lidar,
  long=light detection and ranging
}
\DeclareAcronym{LOS}{
  short=LOS,
  long=line-of-sight
}
\DeclareAcronym{LPF}{
  short=LPF,
  long=low pass filter
}
\DeclareAcronym{LTE}{
  short=LTE,
  long=long term evolution
}
\DeclareAcronym{MIMO}{
  short=MIMO,
  long=multiple-input multiple-output
}
\DeclareAcronym{ML}{
  short=ML,
  long=machine learning
}
\DeclareAcronym{mmWave}{
  short=mmWave,
  long=millimeter wave
}
\DeclareAcronym{MRR}{
  short=MRR,
  long=medium range radar
}
\DeclareAcronym{NLOS}{
  short=NLOS,
  long=non-line-of-sight
}
\DeclareAcronym{NB}{
  short=NB,
  long=narrow beam
}
\DeclareAcronym{NR}{
  short=NR,
  long=new radio
}
\DeclareAcronym{OFDM}{
  short=OFDM,
  long=orthogonal frequency-division multiplexing
}
\DeclareAcronym{ppm}{
  short=ppm,
  long=parts-per-million
}
\DeclareAcronym{PF}{
  short=PF,
  long=particle filter
}
\DeclareAcronym{RMS}{
  short=RMS,
  long=root-mean-square
}
\DeclareAcronym{RPE}{
  short=RPE,
  long=relative precoding efficiency
}
\DeclareAcronym{RS}{
  short=RS,
  long=reference signal
}
\DeclareAcronym{RSRP}{
  short=RSRP,
  long=reference signal received power
}
\DeclareAcronym{RSU}{
  short=RSU,
  long=roadside unit
}
\DeclareAcronym{SCS}{
  short=SCS,
  long=subcarrier spacing
}
\DeclareAcronym{SNR}{
  short=SNR,
  long=signal-to-noise ratio
}
\DeclareAcronym{SRS}{
  short=SRS,
  long=Sounding Reference Signal
}

\DeclareAcronym{SSB}{
  short=SSB,
  long=Synchronization Signal Block
}
\DeclareAcronym{THz}{
  short=THz,
  long=terahertz
}
\DeclareAcronym{UAV}{
  short=UAV,
  long=unmanned aerial vehicle
}
\DeclareAcronym{UE}{
  short=UE,
  long=user equipment
}
\DeclareAcronym{UKF}{
  short=UKF,
  long=unscented Kalman filter
}
\DeclareAcronym{ULA}{
  short=ULA,
  long=uniform linear array
}

\DeclareAcronym{UPA}{
  short=UPA,
  long=uniform planar array
}

\DeclareAcronym{V2I}{
  short=V2I,
  long=vehicle-to-infrastructure
}
\DeclareAcronym{V2V}{
  short=V2V,
  long=vehicle-to-vehicle
}
\DeclareAcronym{V2X}{
  short=V2X,
  long=vehicle-to-everything
}
\DeclareAcronym{VR}{
  short=VR,
  long=virtual reality
}
\DeclareAcronym{VRU}{
  short=VRU,
  long=vulnerable road user
}
\DeclareAcronym{WB}{
  short=WB,
  long=wide beam
}
\DeclareAcronym{BF}{
  short=BF,
  long=beamforming
}
\DeclareAcronym{CS}{
  short=CS,
  long=compressive sensing
}
\DeclareAcronym{NN}{
  short=NN,
  long=neural network
}
\DeclareAcronym{RF}{
  short=RF,
  long=radio frequency
}
\DeclareAcronym{MISO}{
  short=MISO,
  long=multiple-input single-output
}
\DeclareAcronym{SIMO}{
  short=SIMO,
  long=single-input multiple-output
}
\DeclareAcronym{MLP}{
  short=MLP,
  long=multilayer perceptron
}
\DeclareAcronym{AMCF}{
  short=AMCF,
  long=alternative minimization method with a closed-form expression
}
\DeclareAcronym{AWGN}{
  short=AWGN,
  long=additive white Gaussian noise
}
\DeclareAcronym{t-SNE}{
  short=t-SNE,
  long=t-distributed stochastic neighbor embedding
}
\DeclareAcronym{TDD}{
  short=TDD,
  long=time division duplex
}
\DeclareAcronym{mMTC}{
  short=mMTC,
  long=massive Machine Type Communications
}
\DeclareAcronym{NMSE}{
  short=NMSE,
  long=normalized mean square error
}
\DeclareAcronym{MSE}{
  short=MSE,
  long=mean square error
}
\DeclareAcronym{CNN}{
  short=CNN,
  long=convolutional neural network
}
\DeclareAcronym{Tx}{
  short=Tx,
  long=transmit
}
\DeclareAcronym{Rx}{
  short=Rx,
  long=receive
}
\DeclareAcronym{GF}{
  short=GF,
  long=grid-free
}
\DeclareAcronym{CB}{
  short=CB,
  long=codebook-based
}
\DeclareAcronym{DL}{
  short=DL,
  long=deep learning
}
\DeclareAcronym{OOB}{
  short=OOB,
  long=out-of-band
}
\DeclareAcronym{CI}{
  short=CI,
  long=context information
}
\DeclareAcronym{RL}{
  short=RL,
  long=reinforcement learning
}
\DeclareAcronym{MRT}{
  short=MRT,
  long=maximum ratio transmission
}
\DeclareAcronym{MRC}{
  short=MRC,
  long=maximum ratio combining
}
\DeclareAcronym{EGT}{
  short=EGT,
  long=equal gain transmission
}
\DeclareAcronym{EGC}{
  short=EGC,
  long=equal gain combining
}
\DeclareAcronym{DNN}{
  short=DNN,
  long=deep neural network
}
\DeclareAcronym{ReLU}{
  short=ReLU,
  long=rectified linear unit
}
\DeclareAcronym{PSD}{
  short=PSD,
  long=power spectral density
}
\DeclareAcronym{MIB}{
  short=MIB,
  long=Master Information Block
}
\DeclareAcronym{SIB}{
  short=SIB,
  long=System Information Block
}
\DeclareAcronym{PSS}{
  short=PSS,
  long=primary synchronization signal
}
\DeclareAcronym{SSS}{
  short=SSS,
  long=secondary synchronization signal
}

\begin{document}

\title{Grid-Free MIMO Beam Alignment through Site-Specific Deep Learning}

\author{Yuqiang~Heng,
        and~Jeffrey~G.~Andrews,~\IEEEmembership{Fellow,~IEEE}
\thanks{Yuqiang Heng and Jeffrey G. Andrews are with 6G@UT, the Wireless Networking and Communications Group (WNCG), the University of Texas at Austin, Austin, TX 78701 USA. Email: (yuqiang.heng@utexas.edu, jandrews@ece.utexas.edu).   A preliminary version of this work was presented in Globecom 2022 \cite{heng2022GF_globecom}.   
}
}
\maketitle

\begin{abstract}
Beam alignment is a critical bottleneck in \acl*{mmWave} communication. An ideal beam alignment technique should achieve high \acl*{BF} gain with low latency, scale well to systems with higher carrier frequencies, larger antenna arrays and multiple \aclp*{UE}, and not require hard-to-obtain \acl*{CI}. These qualities are collectively lacking in existing methods. We depart from the conventional \acf*{CB} approach where the optimal beam is chosen from quantized codebooks and instead propose a \acl*{GF} beam alignment method that directly synthesizes the \acl*{Tx} and \acl*{Rx} beams from the continuous search space using measurements from a few site-specific probing beams found via a \acl*{DL} pipeline. In realistic settings, the proposed method achieves a far superior \acf*{SNR}-latency trade-off compared to the \acs*{CB} baselines: it aligns near-optimal beams 100x faster or equivalently finds beams with 10-15 dB higher average \acs*{SNR} in the same number of searches, relative to an exhaustive search over a conventional codebook.
\end{abstract}

\IEEEpeerreviewmaketitle

\section{Introduction}\label{section:introduction}
\Ac{mmWave} devices require highly directional \ac{BF} to compensate for the severe isotropic path loss and to achieve viable signal strength. To reduce the cost and power consumption of a fully digital system, practical \ac{mmWave} systems often adopt analog beams that concentrate energy in particular directions. On the other hand, the optimal choice of these narrow beams are susceptible to changes in the propagation environment such as blockage and reflection. It is critical for \ac{mmWave} \acp{BS} and \acp{UE} to find good \ac{BF} directions during initial connection and then track these analog beams as the propagation conditions change, such as when a \ac{UE} moves and rotates. As future cellular systems move to unlock larger bandwidths at higher carrier frequencies extending to the so-called ``sub-\ac{THz}'' bands of up to 300 GHz, devices will adopt ever denser antenna arrays and narrower beams. As a result, beam management -- the process of discovering and maintaining good analog \ac{BF} directions -- is a severe challenge that will only worsen as we move towards 6G and beyond.

\subsection{Background and Related Work} \label{section:related_work}
Existing beam management approaches typically assume a ``grid-of-beams'' or \ac{CB} framework, where \acp{BS} and \acp{UE} adopt codebooks of quantized \ac{BF} directions. In order to ensure coverage in any site, the quantized beams usually distribute energy uniformly in the angular space, such as by using \ac{DFT} codebooks. In this context, the problem of beam alignment becomes selecting the optimal beams from the finite codebooks at the \ac{BS} and potentially at the \ac{UE}. The most widely adopted method of beam selection is through a search. For instance, 5G NR adopts a beam management framework based on beam sweeping, measurement and reporting \cite{Giordani19_BM_Tutorial,heng2021BM_magazine}. In the downlink, the \ac{BS} exhaustively searches its codebook by periodically sending beamformed \acp{RS} called \acp{SSB}. The \ac{UE} measures and reports the received signal strength, and the \ac{BS} then selects the best beam based on the \ac{UE}'s feedback. The obvious drawback of an exhaustive search is its beam sweeping latency, which grows linearly with the total number of beam pairs. As systems move higher in frequency and adopt narrower beams, the size of the codebook increases accordingly. A \ac{THz} system may easily adopt tens of thousands of beam pairs, rendering the the exhaustive search infeasible due to the prohibitive beam sweeping latency. A hierarchical search can reduce the latency for a single \ac{UE}. By sweeping wider beams first and progress to narrower child beams, it iteratively reduces the search space. In the 5G NR framework, the \ac{BS} can use \acp{SSB} to sweep the wide beams and use the more flexible \acp{CSI-RS} for narrow-beam refinement. On the other hand, the radiation patterns of the wide beams need to be carefully designed since the hierarchical search is more prone to search errors caused by noisy measurements and imperfect wide-beam shapes \cite{xiao2016hierarchical,qi2020hierarchical}. 

Another way to reduce the beam sweeping latency is to leverage \acf{CI}. The optimal beam pair of a link is intuitively a function of the topology of the propagation environment and the locations of the \ac{BS} and the \ac{UE}. Such spatial properties can be captured through \ac{CI} such as sub-6 GHz \ac{OOB} measurements \cite{ali2018oob}, \ac{GNSS} coordinates of the \ac{UE} \cite{heng2021MLbeamalignment, va2017inverse}, radar measurements \cite{nuria2016radar,DemirhanAlkhateeb2022RadarBeam}, images captured by cameras and 3D point clouds of the environment \cite{SalehiChowdhury2022multimodal_beam, XuAlkhateeb20203DPointCloudBeam}. By constructing a mapping from the \ac{CI} to the index of the optimal beam through a lookup table or \ac{ML} models, the search space can be substantially reduced. However, these \ac{CI}-based methods are difficult to standardize and are not widely-adopted in practice due to the requirement of additional sensors and the varying hardware capabilities of \ac{mmWave} devices. They are also not suitable for standalone systems due to the need of a robust feedback link usually occupying a lower frequency band.

Inspired by the success of \ac{DL} in computer vision and natural language processing, recent works have explored \ac{DL} in beam management. \Ac{NN} models can extract features from side information such as \ac{CI} \cite{XuAlkhateeb20203DPointCloudBeam,heng2021MLbeamalignment,DemirhanAlkhateeb2022RadarBeam,SalehiChowdhury2022multimodal_beam} and history measurements \cite{Lim2021DL_Kalman_BeamTracking} to predict the optimal beam. \Ac{RL} approaches have been proposed to learn beam management policies through iterative interactions with the environment \cite{mismar20RLbeamforming,xue2022FRLbeam}. Reviews of beam management solutions based on \ac{ML} and \ac{DL} can be found in \cite{ma2022DL_BM_survey,khan2023DL_BM_survey}.

\subsection{Site-specific Adaptation}
A promising way to reduce the beam sweeping latency is through site-specific adaptation. In scenarios where the \ac{AoA} and \ac{AoD} distributions of the channel paths are highly non-uniform, such as in particular environment topologies and when there are several spatial clusters of \acp{UE}, some beams in a uniform codebook may never get used. Leveraging this fact, the \ac{BS} can compute the statistics of the entire codebook after an exhaustive training phase and prioritize the most frequently used beams \cite{khormuji2017statistical_codebook,liu2019statistical_beam_training}. By training using explicit \ac{CSI} \cite{AlrabeiahAlkhateeb2022NNcodebook} or through implicit rewards in a \ac{RL} framework \cite{ZhangAlkhateeb2021RLcodebook}, \acp{NN} can generate smaller codebooks from scratch that adapt to the environment and strategically direct energy towards the \acp{UE}. The data-driven approach can also help select analog beams from standard codebooks using sensing measurements by optimizing the mapping function \cite{ma2020MLbeamalignment} or jointly learning the sensing beams and the mapping function \cite{li19hybrid}. 

Compared to \ac{CB} beam management methods whose \ac{BF} gain is limited by the resolution of the codebook, a \ac{GF} approach extends the search space for the optimal beam to the high-dimensional continuous space and can in theory achieve better gain. For instance, \ac{MRT} is known to be optimal under the unit power constraint and \ac{EGT} under the unit modulus constraint. However, they require full \ac{CSI} to compute the \ac{BF} weights, which is generally unavailable before beam alignment. Furthermore, even with \ac{CSI}, finding the optimal \ac{EGT} and \ac{EGC} beams requires a grid search over possible weights in the \ac{MIMO} setting and is computationally prohibitive \cite{love2003EGT}. Recent works have explored directly predicting the hybrid \ac{BF} weights for the \ac{BS} without full \ac{CSI} by learning sensing matrices \cite{AttiahYu2022DLHybridBF} or a sequence of interactive sensing vectors based on feedback of previous measurements \cite{SohrabiYu2022ActiveSensingDL}.  

In our previous work \cite{heng2022DLCB}, we proposed a \ac{CB} beam alignment method that uses the measurements of a few site-specific probing beams to predict the index of the optimal \ac{Tx} narrow beam. However, it benefits significantly from trying a few candidate beams for each \ac{UE}. As a result, the gain in the beam sweeping overhead diminishes with increasing number of \acp{UE}. This is a common and important limitation of many existing methods: the hierarchical search requires searching all child beams, many \ac{CI}-based models require trying the most likely candidates, and active learning-based methods require sweeping a different sequence of sensing beams for each \ac{UE}. A beam sweeping procedure needs to be repeated for all \acp{UE} in the cell, each of which may require a different set of beams. Hence the total beam sweeping latency increases linearly with the number of \acp{UE}, which can be large and unknown in cellular systems. In \ac{CB} approaches, eventually the entire codebook will need to be searched, a major shortcoming of many of the aforementioned methods.

\subsection{Contributions}

In this work, we propose DL-GF, a one-shot \acf{GF} beam alignment method based on unsupervised \acf{DL}. The proposed method learns a small number of probing beams through site-specific training and directly predicts the optimal analog \ac{BF} vectors without using a quantized codebook. It possesses several qualities of an ideal beam alignment method, many of which are lacking in existing approaches:
\begin{itemize}
    \item \textbf{High \ac{BF} gain beyond standard codebooks.} The proposed method synthesizes continuous-valued analog \ac{BF} weights. With the additional degree of freedom, the synthesized beams can achieve better \ac{BF} gain compared to choosing optimally from large quantized codebooks. 
    \item \textbf{Low beam sweeping latency with optimal scaling.} With site-specific training, the proposed method requires sweeping just a few probing beam pairs to capture sufficient channel information. Unlike many existing approaches that require sweeping a few candidate beams or child beams for each \ac{UE}, the proposed method synthesizes the analog beams in one-shot. As a result, the beam sweeping latency does not increase with the number of \acp{UE}. In our setting, DL-GF reduces the beam sweeping overhead by over 500$\times$ vs. the exhaustive \ac{CB} method regardless of the number of \acp{UE}.
    \item \textbf{Easy to adopt in cellular standards.} The proposed method does not require hard-to-obtain \ac{CI} such as the location of \acp{UE} and \ac{OOB} measurements. Instead, it purely relies on measuring and reporting a few probing beams. The probing beams are also optimized for coverage, allowing the \ac{BS} to discover new \acp{UE}. 
    \item \textbf{Joint \ac{Tx}-\ac{Rx} beam alignment.} Unlike many existing methods that only consider beam alignment for the \ac{BS}, the proposed method jointly predicts the analog beam for both the \ac{BS} and the \ac{UE} without any additional beam sweeping for the \ac{UE}. The synthesized \ac{Tx} and \ac{Rx} beam pairs are jointly optimized to achieve high \ac{BF} gain, even when the \acp{UE} have random orientations.
\end{itemize}

A preliminary version of this work considers \ac{Tx}-only beam alignment \cite{heng2022GF_globecom}. This work extends the conference version to consider \ac{Tx} and \ac{Rx} joint beam alignment in the \ac{MIMO} setting, proposes a more sophisticated utility function, and presents much more comprehensive results.

The rest of this article is organized as follows. The system model is described in Section \ref{section:system_model}. The proposed beam alignment approach, the appropriate metrics and the baselines of comparison are explained in Section \ref{section:proposed_method}. The datasets used are described in Section \ref{section:dataset}. The simulation results are presented in Section \ref{section:evaluation}. We provide the conclusion and final remarks in Section \ref{section:conclusion}.

\textbf{Notation:} 
The following notations are used in this paper: $|a|$ denotes the magnitude of the scalar $a$, $||\mathbf{A}||_{\mathrm{F}}$ denotes the Frobenius norm, $\mathbf{A}^{\mathrm{real}}, \mathbf{A}^{\mathrm{imag}}$ denote the real and imaginary parts of a complex matrix $\mathbf{A}$, $\mathbf{A}^{T}$ denotes the transpose, $\mathbf{A}^{H}$ denotes the conjugate transpose, $\mathrm{diag}(\mathbf{a})$ denotes the diagonal matrix constructed from the vector $\mathbf{a}$, $\mathrm{diag}(\mathbf{A})$ denotes the vector of diagonal elements of the matrix $\mathbf{A}$, $[\mathbf{a}]_i$ denotes the $i$th element of the vector $\mathbf{a}$, $\mathbf{a} \otimes \mathbf{b}$ denotes the Kronecker product, $\mathbf{A} \oslash \mathbf{B}$ denotes the element-wise division and $\mathbf{A}^{|\cdot|}$ denotes the element-wise magnitude.

\section{System Model}\label{section:system_model}
A \ac{MIMO} system is considered where the \ac{BS} has an array of $N_{\textnormal{T}}$ antennas, the \ac{UE} has an array of $N_{\textnormal{R}}$ antennas and both perform beam alignment. A geometric channel model with $L$ paths is adopted:
\begin{equation}
    \mathbf{H} = \sum_{l=1}^{L} \alpha_{l} e^{j(\theta_{l}-2\pi\tau_{l}B)} \mathbf{a}_{\textnormal{R}}(\omega_{l}^{az},\omega_{l}^{el}) \mathbf{a}_{\textnormal{T}}(\phi_{l}^{az},\phi_{l}^{el})^{H},
\end{equation}
where $\mathbf{a}_{\textnormal{T}}, \mathbf{a}_{\textnormal{R}}$ are the \ac{Tx} and \ac{Rx} array response vectors, $\alpha_l, \theta_l, \tau_l$ are the gain, Doppler shift and delay, $\omega_{l}^{az}, \omega_{l}^{el}, \phi_{l}^{az}, \phi_{l}^{el}$ are the azimuth and elevation \ac{AoA} and \ac{AoD} of path $l$. 
For a \ac{UPA} with $N_y$ and $N_z$ antennas in the $y-z$ plane, its array response vector is given by:
\begin{multline}
    \mathbf{a}_y(\phi^{az},\phi^{el}) = \\
\begin{bmatrix}
1 & e^{j\frac{2 \pi}{\lambda}d\sin \phi^{az} \sin \phi^{el}} & \cdots & e^{j(N_{y}-1)\frac{2 \pi}{\lambda}d\sin \phi^{az} \sin \phi^{el}}
\end{bmatrix}^{T},
\end{multline}
\begin{equation}
    \mathbf{a}_z(\phi^{el}) = 
\begin{bmatrix}
1 & e^{j\frac{2 \pi}{\lambda}d \cos \phi^{el}} & \cdots & e^{j(N_{z}-1)\frac{2 \pi}{\lambda}d \cos \phi^{el}}
\end{bmatrix}^{T},
\end{equation}
$\lambda$ is the carrier wavelength and $d$ is the antenna spacing.

During data transmission in the downlink, if the \ac{BS} adopts a \ac{Tx} beam $\mathbf{v}_{\textnormal{T}} \in \mathbbm{C}^{N_{\textnormal{T}} \times 1}$ and transmits a symbol $s$ and the \ac{UE} adopts a \ac{Rx} beam $\mathbf{v}_{\textnormal{R}} \in \mathbbm{C}^{N_{\textnormal{R}} \times 1}$, the received signal can be written as
\begin{equation}
    y = \sqrt{P_{\textnormal{T}}}\mathbf{v}_{\textnormal{R}}^{H}\mathbf{H}\mathbf{v}_{\textnormal{T}}s + \mathbf{v}_{\textnormal{R}}^{H}\mathbf{n},
\end{equation}
where $P_{\textnormal{T}}$ is the \ac{Tx} power and $\mathbf{n} \sim \mathcal{CN}(0,\sigma^2\mathbb{I})$ is a complex \ac{AWGN}. Assuming unit-power transmitted symbols, the \ac{SNR} achieved by the beam pair is 
\begin{equation}\label{eq:data_snr}
    \textnormal{SNR} = \frac{P_{\textnormal{T}}|\mathbf{v}_{\textnormal{R}}^{H}\mathbf{H}\mathbf{v}_{\textnormal{T}}|^2}{\sigma^2}.
\end{equation}

The \ac{BS} and the \acp{UE} are assumed to perform analog \ac{BF} only. Each device has a single \ac{RF} chain connected to an array of phase shifters. Hence the \ac{BF} vectors satisfy the unit-power, constant modulus constraint:
\begin{equation}
|[\mathbf{v}_{\textnormal{T}}]_i| = \frac{1}{\sqrt{N_{\textnormal{T}}}}, i=1, \dots, N_{\textnormal{T}} \
\end{equation}
\begin{equation}
|[\mathbf{v}_{\textnormal{R}}]_j| = \frac{1}{\sqrt{N_{\textnormal{R}}}}, j=1, \dots, N_{\textnormal{R}}.
\end{equation}

The orientation of a \ac{UE} is modeled with random rotations with respect to the local coordinate system which is attached to and rotates with the \ac{UE} \cite{3gpp.NR.Channel}. Each \ac{UE} first rotates by the $z$ axis, then by the new $y$ axis, and finally by the new $x$ axis. The angles of the three elemental rotations are modeled as random variables. The \ac{AoA} of the channel is then adjusted according to the orientation of the \ac{UE}. A more sophisticated model may consider the shape of the \ac{UE} and self--blockage of the device, which is left for future work.

\section{The Proposed Method, Metrics and Baselines}\label{section:proposed_method}
Our objective is to directly predict optimal \ac{Tx} and \ac{Rx} beams without searching large codebooks or candidate beams. In our previous work \cite{heng2022DLCB}, we demonstrated that a \ac{NN} classifier can accurately select the optimal beam index from a large codebook using measurements of a few learned site-specific probing beams, i.e., each \ac{BS} learns unique probing beams that are well-suited to its propagation environment. Inspired by this idea, we propose to learn a small number of probing beam pairs at the \ac{BS} and the \ac{UE} then use their measurements to synthesize the optimal narrow beam pair from the continuous search space. The \ac{BS} first sweeps $N_{\mathbf{F}}$ \ac{Tx} probing beams while the \ac{UE} measures the received signal power using $N_{\mathbf{W}}$ \ac{Rx} sensing beams. After the probing-beam sweeping phase, the \ac{UE} uses the collected measurements as inputs to its beam synthesizer function $f_{\textnormal{R}}$ to generate its \ac{Rx} beam. The measurements are also fed back to the \ac{BS}, which uses them as inputs to its own beam synthesizer function $f_{\textnormal{T}}$ to generate the \ac{Tx} beam. 

\begin{figure}[!tbp]
  \centering
  \includegraphics[width=0.6\textwidth]{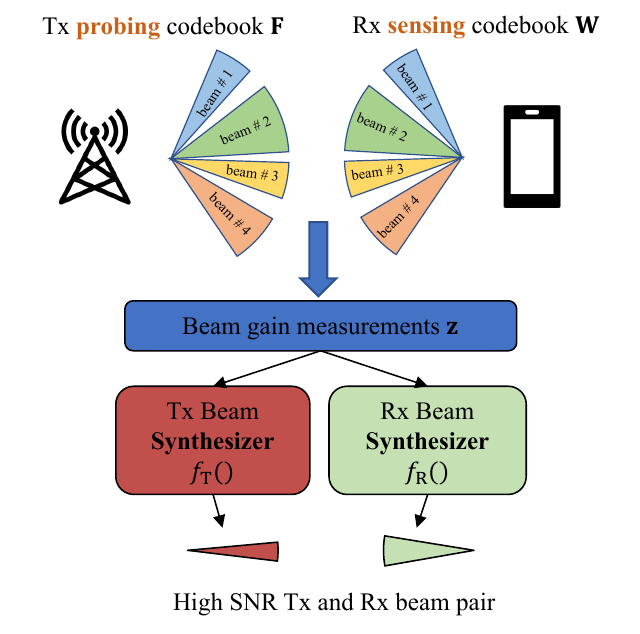}
  \caption{Illustration of the proposed beam alignment procedure based on site-specific probing. The learned probing and sensing beams are implemented in \ac{RF} and measured through beam sweeping.}\label{figure:deployment_illr}
\end{figure}

The proposed beam alignment procedure is illustrated in Fig. \ref{figure:deployment_illr}.
During the first phase of the proposed method, the \ac{BS} and the \acp{UE} sweep a small number of probing and sensing beams to capture information about the channel.
Let $\mathbf{F} \in \mathbbm{C}^{N_{\textnormal{T}} \times N_{\mathbf{F}}}, \mathbf{W} \in \mathbbm{C}^{N_{\textnormal{R}} \times N_{\mathbf{W}}}$ be matrices representing the \ac{BS} probing and \ac{UE} sensing beams, whose entries satisfy the constant modulus constraint for analog \ac{BF}. 
The \ac{BS} periodically broadcasts probing symbols $\mathbf{s}_{\textnormal{probe}}$ using beams in $\mathbf{F}$ while each \ac{UE} measures using the sensing beams in $\mathbf{W}$. Note that all \acp{UE} in the cell share the same sensing beams. The periodic sweeping of the probing beams is similar to \ac{SSB} transmission in 5G, which also allows new \acp{UE} to be discovered and synchronize to the \ac{BS}. The total number of beams swept does not increase with the number of \acp{UE} since multiple \acp{UE} can measure the probing beams using shared time and frequency resources.

After beam sweeping, the composite received signal $\mathbf{Y} \in \mathbbm{C}^{N_{\mathbf{W}} \times N_{\mathbf{F}}}$ consisting of the received signal of all combinations of probing and sensing beams can be written as
\begin{equation}\label{eq:probing_rx_signal_matrix}
    \mathbf{Y} = \sqrt{P_{\textnormal{T}}}\mathbf{W}^{H}\mathbf{H}\mathbf{F}\mathrm{diag}(\mathbf{s}_{\textnormal{probe}})+\mathbf{W}^{H}\mathbf{N},
\end{equation}
where $\mathbf{s}_{\textnormal{probe}} \in \mathbbm{C}^{N_{\textnormal{F}} \times 1}$ is the vector of transmitted probing symbols and $\mathbf{N} \in \mathbbm{C}^{N_{\textnormal{R}} \times N_{\mathbf{F}}}$ is the \ac{AWGN} measurement noise matrix whose each element has noise power $\sigma_{\textnormal{probe}}^2$. 

The probing beam pairs will likely have lower gain since there are many fewer of them to cover the angular space. To combat the low \ac{SNR} during the probing measurement phase, the \ac{BS} can introduce redundancy in the probing symbols over time or frequency, such as by using a spreading sequence or repeating over subcarriers \cite{Giordani19_BM_Tutorial, Barati2015_cell_discovery}. This is not too wasteful in the context of analog \ac{BF} and 5G beam alignment, since a single analog beam can be transmitted at a time (such as for each \ac{SSB}). A spreading gain of $\beta \geq 1$ is introduced to model the effective \ac{SNR} of the probing measurements:

\begin{equation}\label{eq:snr_enhancement}
  \textnormal{SNR}_{\textnormal{probe}} = \beta\frac{P_{\textnormal{T}}|\mathbf{w}^{H}\mathbf{H}\mathbf{f}|^2}{\sigma^2}.
\end{equation}

We focus on the special case where $N_{\mathbf{F}}=N_{\mathbf{W}}=N_{\mathrm{probe}}$: the \ac{BS} sweeps $N_{\mathrm{probe}}$ probing beams while the \ac{UE} uses a different \ac{Rx} sensing beam for each \ac{Tx} probing beam so that the total number of probing beam pairs is $N_{\mathrm{probe}}$. The \ac{UE} then reports the $N_{\mathrm{probe}}$ received signal power measurements. As a result, the input feature to the beam synthesizer functions is 
\begin{equation}
\mathbf{z} =
\begin{bmatrix}
|[\mathrm{diag}(\mathbf{Y})]_1|^2 & \cdots & |[\mathrm{diag}(\mathbf{Y})]_{N_{\mathrm{probe}}}|^2
\end{bmatrix}^T.
\end{equation}
The beam synthesizer functions $f_{\textnormal{T}},f_{\textnormal{R}}$ at the \ac{BS} and the \ac{UE} then predict the \ac{Tx} and \ac{Rx} beams for data transmission.

In principle, both the measurement and the feedback procedures are flexible and can be design choices. For example, the number of probing and sensing beams can be arbitrary. The \ac{UE} can measure and feedback all combinations of probing and sensing beams, and the beam synthesizer functions can utilize the full measurement matrix $\mathbf{Y} \in \mathbbm{C}^{N_{\mathbf{W}} \times N_{\mathbf{F}}}$ to generate the \ac{BF} vectors. The \ac{UE} may also feedback just $N_{\mathbf{F}}$ measurements corresponding to the strongest sensing beam for each \ac{Tx} probing beam. The specific design considered in this work is motivated by our wish to reduce the measurement and feedback overhead, i.e., $N_{\mathrm{probe}} \ll N_{\mathbf{F}}N_{\mathbf{W}}$. Furthermore, it is also more compatible with the \ac{RS} feedback procedure in 5G NR, where the \ac{UE} beam is transparent to the \ac{BS}. A comparison of different probing strategies is discussed in Sect. \ref{section:ablation_probing_strategy}. However, there is extensive scope for future work exploring other approaches.

\subsection{Problem Formulation}\label{section:problem_formulation}

The parameters of the proposed method include the probing and sensing beams $\mathbf{F},\mathbf{W}$ as well as the beam synthesizers $f_{\textnormal{T}},f_{\textnormal{R}}$, which need to be optimized with respect to a utility function. With the unit modulus constraint of the probing and predicted beam pairs in mind, the optimization problem can be written as:
\begin{equation}\label{eq:optimization_formulation}
\begin{array}{rrcll}
    \displaystyle \max_{\mathbf{F},\mathbf{W},f_{\textnormal{T}},f_{\textnormal{R}}} & & \mathcal{U} & &  \\
    \textrm{s.t.} & \mathbf{v}_{\mathrm{T}} & = & f_{\textnormal{T}}(\mathbf{z}) & \\
    & \mathbf{v}_{\mathrm{R}} & = & f_{\textnormal{R}}(\mathbf{z}) & \\
    &|[\mathbf{v}_{\mathrm{T}}]_{i}| & = & \frac{1}{\sqrt{N_{\textnormal{T}}}}, & \forall i=1,\cdots,N_{\textnormal{T}} \\
    &|[\mathbf{v}_{\mathrm{R}}]_{i}| & = & \frac{1}{\sqrt{N_{\textnormal{R}}}}, & \forall i=1,\cdots,N_{\textnormal{R}} \\
    &|[\mathbf{F}]_{i,j}| & = & \frac{1}{\sqrt{N_{\textnormal{T}}}}, & \forall i=1,\cdots,N_{\textnormal{T}},  \\
    & & & &\forall j=1,\cdots,N_{\mathbf{F}} \\
    &|[\mathbf{W}]_{i,j}| & = & \frac{1}{\sqrt{N_{\textnormal{R}}}}, & \forall i=1,\cdots,N_{\textnormal{R}},  \\
    & & & &\forall j=1,\cdots,N_{\mathbf{W}}. \\
\end{array}
\end{equation}

There are several considerations when designing the utility function. The beam synthesizer functions should generate beams that tend to maximize the \ac{BF} gain for each channel realization. The probing and sensing beams in $\mathbf{F}$ and $\mathbf{W}$ serve two important purposes. First, they provide helpful information to the beam synthesizers and thus should capture characteristics of the channel. Second, they should allow the \ac{BS} to discover new \acp{UE} during the \ac{IA} process and thus should satisfy a minimum \ac{SNR} requirement. Since the optimal \ac{BF} vectors can be easily computed in closed-form in the simpler \ac{MISO} or \ac{SIMO} setting, the synthesized beams can be optimized to resemble the optimal \ac{EGT} or \ac{EGC} beams by minimizing a supervised loss function such as the \ac{MSE} in \cite{AlrabeiahAlkhateeb2022NNcodebook}. However, finding the \ac{EGT} and \ac{EGC} beam pairs in the \ac{MIMO} setting requires solving its own non-convex optimization problem and may require a grid-search over possible weights, making a supervised utility function undesirable. To this end, we propose a two-component unsupervised utility function that does not require explicit labels for the probing or synthesized beams:

\begin{align} \label{eq:bf_loss}
  \mathcal{U} &=  \gamma\mathcal{U}_{\textnormal{BF}} + (1-\gamma)\mathcal{U}_{\textnormal{IA}} \\
  \mathcal{U}_{\textnormal{BF}} &= \mathop{\mathbb{E}}\limits_{\mathbf{H} \in \mathcal{H}} \left[ \frac{|\mathbf{v}_\mathrm{R}^H \mathbf{H} \mathbf{v}_\mathrm{T}|^2}{||\mathbf{H}||^2_{\text{F}}} \right] \\
  \mathcal{U}_{\textnormal{IA}} &= \mathop{\mathbb{E}}\limits_{\mathbf{H} \in \mathcal{H} \setminus \mathcal{H}_{\textnormal{IA}}} \left[ \max\limits_{i \in {1,\cdots,N_{\textnormal{probe}}}} \frac{|\mathbf{w}_{i}^H \mathbf{H} \mathbf{f}_{i}|^2}{||\mathbf{H}||^2_{\text{F}}} \right] \\
  \mathcal{H}_{\textnormal{IA}} &= \{\mathbf{H} \in \mathcal{H}: \max\limits_{i \in {1,\cdots,N_{\textnormal{probe}}}} \frac{P_{\textnormal{T}}|\mathbf{w}_{i}^H \mathbf{H} \mathbf{f}_{i}|^2}{\sigma^2} \geq \textnormal{SNR}_{\textnormal{TH}} \}
\end{align}

Let $\mathcal{H}$ denote the set of channel realizations corresponding to possible \ac{UE} locations in the cell. The first term $\mathcal{U}_{\textnormal{BF}}$ focuses on the \ac{BF} gain of the synthesized beams $\mathbf{v}_\mathrm{T},\mathbf{v}_\mathrm{R}$ and is the end-to-end objective. An obvious choice for the utility function is the average \ac{SNR} or achievable rate of the predicted beam pairs, such as adopted in \cite{AttiahYu2022DLHybridBF}. However, it tends to emphasize \acp{UE} with good channels more and neglect cell edge \acp{UE}. In order to provide better coverage to all \acp{UE}, we maximize the average \ac{BF} gain normalized by the channel norm to give equal emphasis even for \acp{UE} with worse channels. It also performs better than simply maximizing the average rate.

The second term $\mathcal{U}_{\textnormal{IA}}$ ensures the coverage of the probing beams, which are transmitted using \acp{SSB} in 5G so that a new \ac{UE} may synchronize and connect with the \ac{BS}. The strongest probing beam pair for a \ac{UE} needs to exceed an \ac{SNR} threshold $\textnormal{SNR}_{\textnormal{TH}}$ for that \ac{UE} to be discovered.
In the context of 5G, a new \ac{UE} needs to decode the synchronization signals, \ac{MIB} and \ac{SIB} in an \ac{SSB}, which may not enjoy the proposed spreading gain of the probing symbols.
Let $\mathbf{f}_i$ and $\mathbf{w}_i$ denote the $i$th probing and sensing beams, the \ac{SNR} for \ac{IA} is that of the strongest probing beam pair:
\begin{equation}\label{eq:snr_IA}
  \textnormal{SNR}_{\textnormal{IA}} = \max\limits_{i \in {1,\cdots,N_{\textnormal{probe}}}} \frac{P_{\textnormal{T}}|\mathbf{w}_{i}^H \mathbf{H} \mathbf{f}_{i}|^2}{\sigma^2}.
\end{equation}
Let $\mathcal{H}_{\textnormal{IA}}$ denote the set of channel realizations corresponding to \acp{UE} that can achieve the \ac{SNR} threshold $\textnormal{SNR}_{\textnormal{TH}}$ for \ac{IA} with their strongest probing beam pairs. The second term $\mathcal{U}_{\textnormal{IA}}$ seeks to maximize the average normalized \ac{BF} gain of the strongest probing beam pairs for those \acp{UE} that cannot meet the minimum \ac{SNR} threshold $\textnormal{SNR}_{\textnormal{TH}}$ with any of the probing beam pairs. The coefficient $\gamma \in [0,1]$ is a design parameter to balance the trade-off between the gain of the synthesized beams and of the probing beam pairs.

We focus on the beam alignment problem that finds analog beams for each \ac{UE} individually. Systems that adopt hybrid \ac{BF} can first perform beam alignment and select good analog beams, then design digital beams that optimize for inter-user interference and sum-rate based on the effective channel. Joint optimization of analog and digital beams in a multi-\ac{UE} setting is a promising direction and is left for future work.

\subsection{The Proposed NN Architecture}\label{section:NN_architecture}

\begin{figure}[!tbp]
  \centering
  \includegraphics[width=0.8\textwidth]{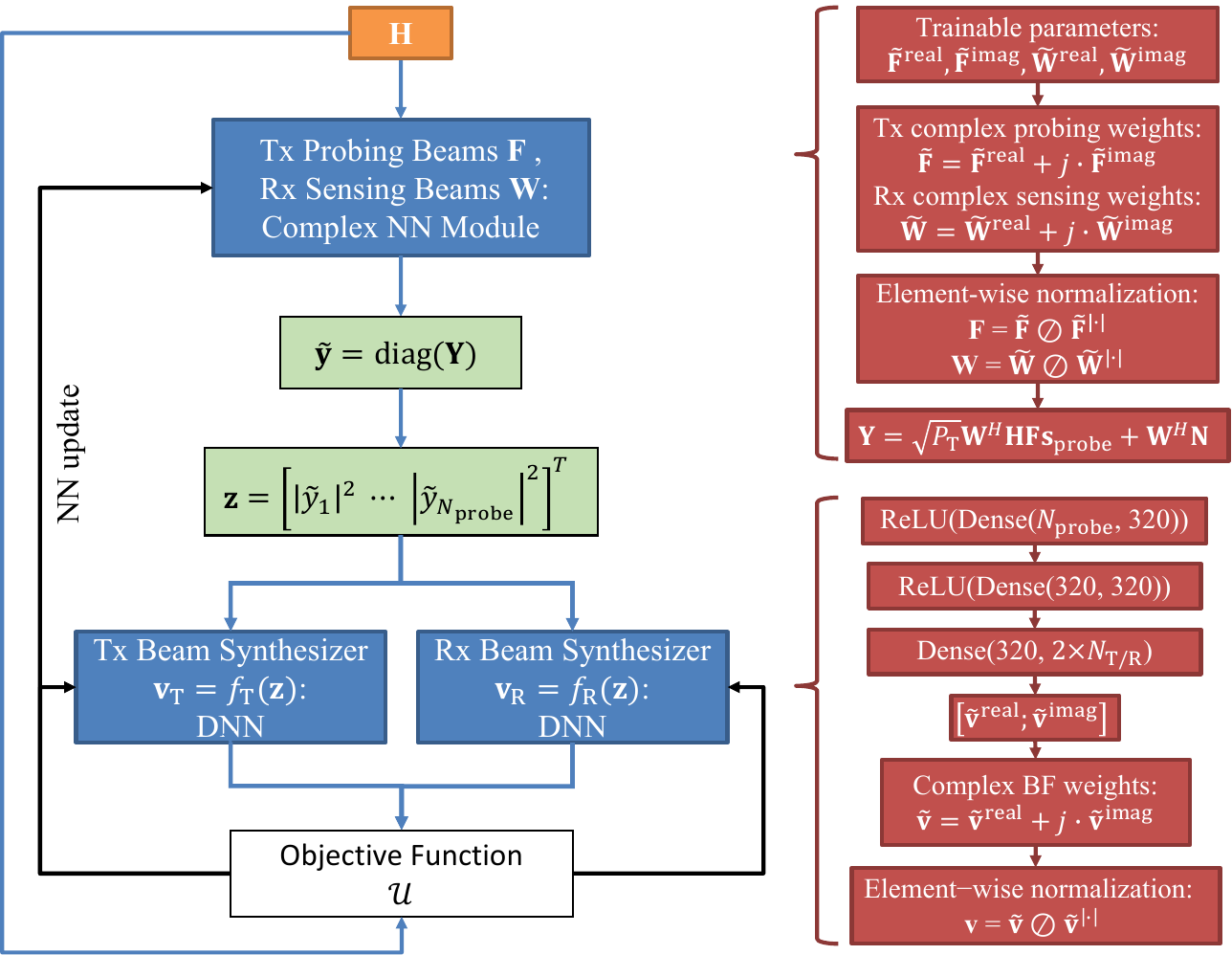}
  \caption{The architecture of the proposed NN, including the probing beam pairs $\mathbf{F}$ and $\mathbf{W}$ and the beam selection functions $f_{\textnormal{T}}(\cdot)$ and $f_{\textnormal{R}}(\cdot)$. The \ac{Tx} and \ac{Rx} beam synthesizers have the same architecture.}\label{figure:NN_architecture_training}
\end{figure}

The optimization problem in (\ref{eq:optimization_formulation}) is difficult to solve since the unit-modulus constraints are non-convex while the functions $f_{\textnormal{T}}$ and $f_{\textnormal{R}}$ are generally unknown. We propose to parameterize the probing and sensing beams $\mathbf{F},\mathbf{W}$ as well as the beam synthesizers $f_{\textnormal{T}},f_{\textnormal{R}}$ using \acp{NN} and optimize them in a data-driven fashion. The overall \ac{NN} architecture is illustrated in Fig. \ref{figure:NN_architecture_training}. During the offline training phase, the input to the entire \ac{NN} model are the channel matrices. The \ac{Tx} probing and \ac{Rx} sensing beams are implemented in the complex \ac{NN} module as two complex matrices. The entire complex \ac{NN} module essential computes complex-valued matrix multiplication and addition on the channel matrix. To enforce the unit-modulus constraint, the complex matrices are normalized element-wise by the magnitude. We find this implementation to perform better empirically compared to the implementation used in our previous work \cite{heng2022DLCB}, where the complex matrices are computed using the phase-shift values. The complex \ac{NN} module computes the composite matrix of received signals of all combinations of probing and sensing beams in (\ref{eq:probing_rx_signal_matrix}) and can be considered to perform virtual sweeping of the probing beam pairs while being differentiable.
The power of the $N_{\textnormal{probe}}$ probing beam pairs corresponding to the diagonal elements are extracted and fed into the \ac{Tx} and \ac{Rx} beam synthesizer functions, each parameterized using an \ac{MLP}. The \ac{MLP} consists of 2 hidden layers with \ac{ReLU} activation and a final linear layer outputting the real and imaginary parts of the synthesized \ac{BF} vector. The final predicted beams are normalized element-wise to enforce the unit-modulus constraint. With each batch of training channel realizations, the utility can be computed using the \ac{BF} gain of the synthesized beams and that of the best probing beam pairs. The probing and sensing beams as well as the beam synthesizer functions are trained through stochastic gradient descent and backpropagation. 

\subsection{Practicality of the Proposed Method}
The \ac{NN} model is trained offline prior to deployment. The training data can be obtained through ray-tracing simulations or from measurements. Due to the difficulty of obtaining \ac{CSI} in real time, online training and adaptation of the \ac{NN} remain an open research problem. On the other hand, the model proves to be fairly robust against imperfect training data, as we will show in Section \ref{section:imperfect_training_data}. After the offline training phase, the probing beam pairs are extracted from the complex \ac{NN} module and implemented in \ac{RF}. The beam synthesizers $f_{\mathrm{T}}, f_{\mathrm{R}}$ are implemented at the \ac{BS} and the \ac{UE} respectively. During deployment, the \ac{BS} and the \ac{UE} periodically sweep the probing beam pairs while the \ac{UE} measures the received signal power. The probing measurements are then reported to the \ac{BS}. The \ac{BS} and the \ac{UE} use the probing measurements as inputs to $f_{\mathrm{T}}$ and $f_{\mathrm{R}}$ to synthesize the \ac{Tx} and \ac{Rx} beams. Since the probing beams and the beam synthesizers are site-specific, $\mathbf{W}$ and $f_{\mathrm{R}}$ need to be transmitted to the \acp{UE}. In a non-standalone system, this can be done through a lower-frequency side link. A new \ac{UE} can then complete \ac{IA} using the strongest probing beam pair. We have previously shown that the \ac{Tx} probing beams can be used for \ac{IA} in the \ac{MISO} setting \cite{heng2022GF_globecom}. Therefore in a standalone system, the \ac{BS} may sweep the \ac{Tx} probing beams while a new \ac{UE} receives using a quasi-omnidirectional beam. The \ac{UE} can complete \ac{IA} with the strongest \ac{Tx} probing beam, download the \ac{Rx} components $\mathbf{W}$ and $f_{\mathrm{R}}$, then perform the full joint \ac{Tx}-\ac{Rx} beam alignment.

\subsection{Baselines and Metrics}
The proposed method optimizes the \ac{GF} beam synthesizers and the probing beam pairs through \ac{DL}, hence is referred to as the DL-GF method. It will be compared with five baselines: the \ac{DL}-\ac{CB} method, the exhaustive search, the genie with DFT codebooks, the DFT+\ac{EGC} method, and \ac{MRT}+\ac{MRC}. The same probing measurement spreading gain of $\beta$ is adopted in DL-GF, DL-CB and the exhaustive search for a fair comparison.
\begin{itemize}
    \item \textbf{DL-CB.} The DL-CB method replaces the \ac{Tx} and \ac{Rx} beam synthesizers with two classifiers that predict the optimal beam indices in the \ac{BS} and \ac{UE} codebooks. It is an extension of our previous work \cite{heng2022DLCB} to the \ac{MIMO} scenario. To improve its accuracy, the DL-CB method can sweep the top few candidate beam pairs predicted by the classifiers. In our experiment, the best beam pair is selected after trying all combinations of the top-3 predicted \ac{Tx} and \ac{Rx} beams.
    \item \textbf{Exhaustive search.} The exhaustive search sweeps and measures all combinations of beam pairs in the \ac{BS} and \ac{UE} codebooks and selects the beam pair with the highest received signal power. Since the measurements are corrupted by noise, the beam pair selected by the exhaustive search may not be the one maximizing \ac{SNR}. 
    \item \textbf{Genie DFT.} A genie always selects the optimal beam pair in the \ac{BS} and \ac{UE} codebooks. It is the same as the exhaustive search when the measurement noise power is zero.
    \item \textbf{DFT+EGC.} Only the \ac{BS} has a codebook in the DFT+EGC baseline. The \ac{BS} exhaustively tries all beams in its codebook while the \ac{UE} uses the corresponding \ac{EGC} vector for each \ac{Tx} beam. The best beam pair is selected assuming no measurement noise. It is expected to perform better than the genie DFT baseline due to the additional degree of freedom at the \ac{UE}.
    \item \textbf{MRT+MRC.} Neither the \ac{BS} nor the \ac{UE} uses a codebook. The \ac{BS} uses the optimal \ac{MRT} beam while the \ac{UE} uses the \ac{MRC} beam. The \ac{BF} gain can be computed through an eigendecomposition of $\mathbf{H}^H\mathbf{H}$. The MRT+MRC baseline is the theoretical upper bound under the unit-power constraint and cannot be achieved under the stricter unit-modulus constraint.
\end{itemize}

One important performance metric for beam alignment is the \ac{SNR} achieved by the selected beams. We will compare the average \ac{SNR} as well as the \ac{SNR} distribution across different baselines. Secondly, new \acp{UE} need to satisfy a minimum \ac{SNR} requirement so that they can be discovered during the \ac{IA} process and connect to the \ac{BS}. 
This is generally not a concern for \ac{CB} approaches that adopt uniform codebooks, but may be problematic when the probing beams are site-specific and have severe coverage holes. Therefore, we will also investigate the misdetection probability of \acp{UE}, which is the probability that a \ac{UE} achieves below-threshold \ac{SNR} with the strongest probing beam pair, formally defined as
\begin{equation}
 P\Bigl(\max\limits_{i \in {1,\cdots,N_{\textnormal{probe}}}} \frac{P_{\textnormal{T}}|\mathbf{w}_{i}^H \mathbf{H} \mathbf{f}_{i}|^2}{\sigma^2} < \textnormal{SNR}_{\textnormal{TH}}\Bigr).
\end{equation}
 The \ac{SNR} threshold is chosen to be -5 dB \cite{giordani16initial_access}.

\section{Dataset}\label{section:dataset}
Four scenarios from the public DeepMIMO dataset \cite{alkhateeb2019deepmimo} are considered to capture a wide range of propagation environments, including indoor and outdoor environments, 28 GHz and 60 GHz carrier frequencies, as well as \ac{LOS} and \ac{NLOS} \acp{UE}. The channel realizations are computed through ray-tracing with a state-of-the-art commercial-grade software \cite{Remcom01}, which is one of the most accurate ways of simulating \ac{mmWave} channels once the environment topology is specified. The \ac{BS} and \acp{UE} adopt \acp{UPA} with half-wavelength spacing. The simulation parameters are summarized in Table \ref{table:data_params} for the four scenarios described below.

\begin{table}
\small
\centering
  \caption{Simulation Parameters}\label{table:data_params}
    \begin{tabular}{| c | c |}
    \hline
    BS Antenna & $8\times8$ UPA\\ \hline
    BS Codebook Size & $16\times16=256$\\ \hline
    UE Antenna & $4\times4$ UPA\\ \hline
    UE Codebook Size & $8\times8=64$\\ \hline
    Antenna Element & Isotropic \\ \hline
    Carrier Frequency & \makecell{Outdoor LOS (O1): 28 GHz \\Indoor (I3): 60 GHz \\Outdoor NLOS (O1 blockage): 28 GHz}\\ \hline
    Bandwidth ($B$) & 100 MHz\\ \hline
    Transmit Power ($P_{T}$) & \makecell{O1: 20 dBm \\I3, BS 1: 10 dBm \\I3, BS 2: 20 dBm\\ O1 blockage: 35 dBm}\\ \hline
    Noise Power $\sigma^2$ & -81 dBm \\ \hline
    \makecell{Probing Measurement \\Spreading Gain $\beta$} & 16 \\ \hline
    UE Orientation Range & \makecell{$z:Unif(-\pi,\pi)$ \\$y:Unif(-\frac{\pi}{2},\frac{\pi}{2})$\\$x:Unif(-\frac{\pi}{2},\frac{\pi}{2})$} \\ \hline
    \end{tabular}
\end{table}

\begin{figure}%
\centering
\subfloat[O1]{\includegraphics[width=0.49\columnwidth]{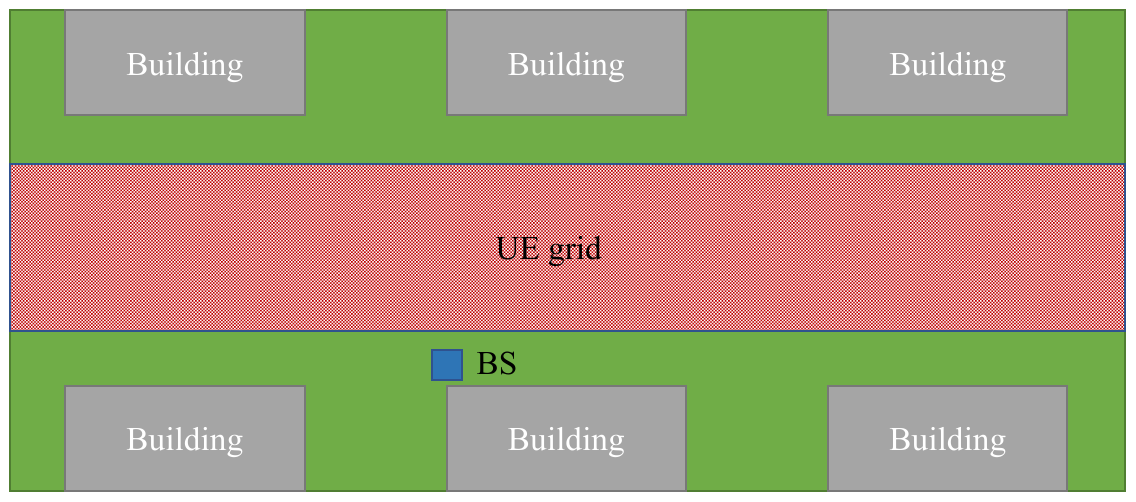}\label{figure:O1_2D}}
\subfloat[O1 blockage]{\includegraphics[width=0.49\columnwidth]{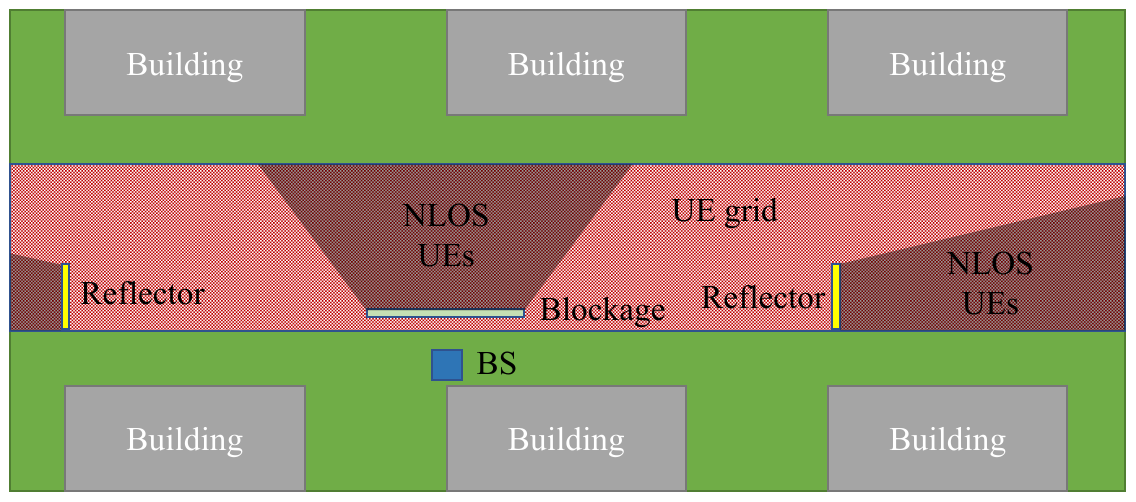}\label{figure:O1B_2D}}
\hfill
\subfloat[I3]{\includegraphics[width=0.45\columnwidth]{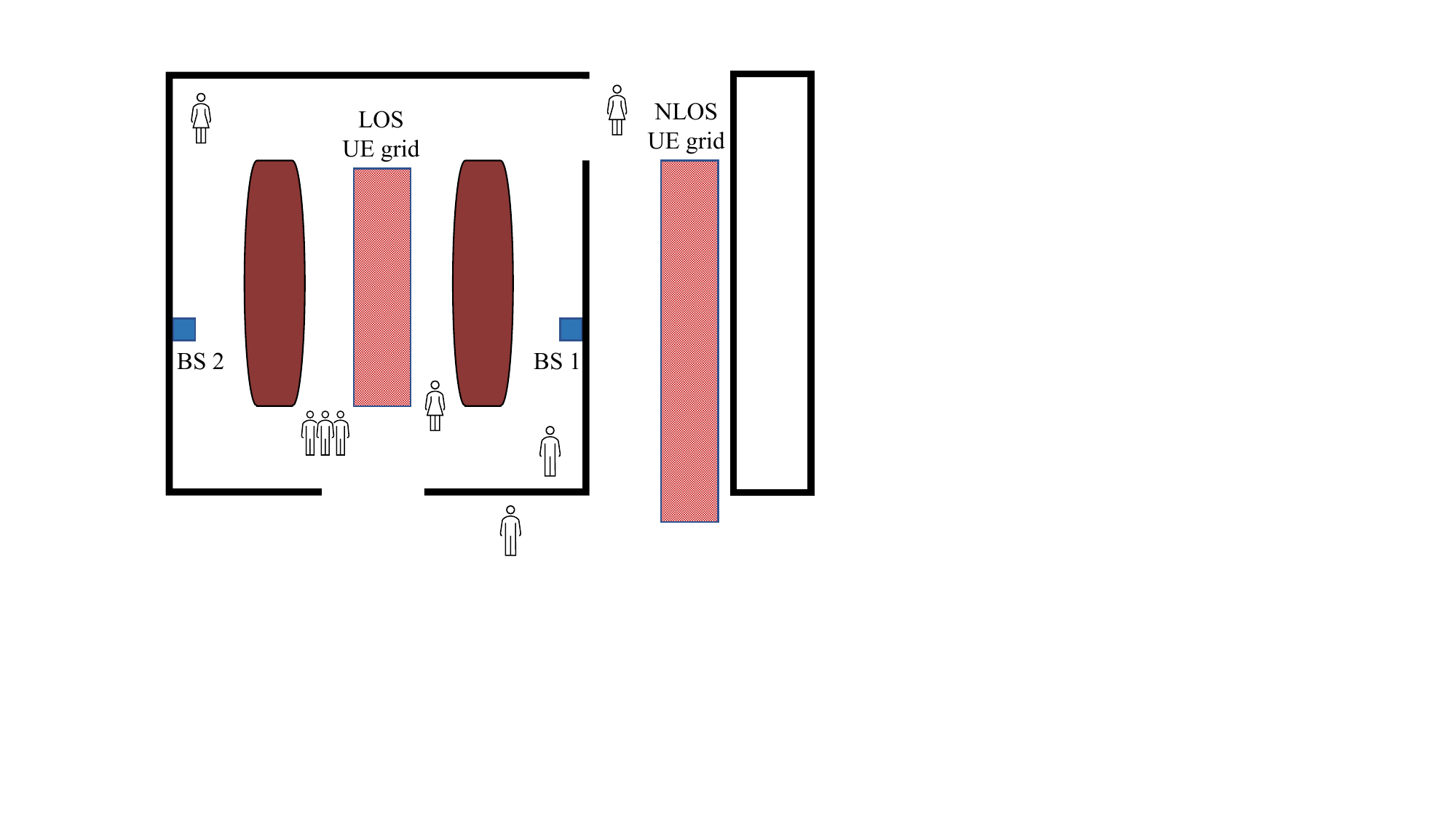}\label{figure:I3_2D}}
\caption{Illustration of the DeepMIMO ray-tracing scenarios.}
\end{figure}

\textbf{O1 Scenario.} The O1 scenario captures an outdoor urban street environment with \ac{LOS} \acp{UE}. An illustration of the scenario is shown in Fig. \ref{figure:O1_2D}. We select \ac{BS} 3 and \acp{UE} from row \#800 to row \#1200 from the original dataset. The \ac{BS} is placed on the street side and a total of 72,581 \acp{UE} are placed on a uniform grid on the street. The carrier frequency is 28 GHz. 

\textbf{I3 Scenarios.} The I3 scenarios captures an indoor office environment with both \ac{LOS} and \ac{NLOS} \acp{UE}. An illustration of the environment is shown in Fig. \ref{figure:I3_2D}. There are two \acp{BS} in this scenario: BS 1 is placed on the inside wall of the conference room and BS 2 is placed on the opposite wall. A grid of \ac{LOS} \acp{UE} is placed inside the conference room and a grid of \ac{NLOS} \acp{UE} is placed in the corridor outside. There are a total of 118,959 \acp{UE} and the carrier frequency is 60 GHz. We consider two scenarios based on this environment, each with one of the two \acp{BS} activated.

\textbf{O1 Blockage Scenario.} The O1 blockage scenario is artificially created based on the O1 scenario with a metal screen placed in front of the \ac{BS} and two reflectors placed on both ends of the street, as illustrated in Fig. \ref{figure:O1B_2D}. The carrier frequency is 28 GHz and there are a total of 497,931 \ac{UE} positions. While it may not be representative of practical \ac{mmWave} deployments, the extreme topology allows us to gain some intuition on the beam patterns learned in this environment.


\section{Evaluation}\label{section:evaluation}
The datasets discussed in Section \ref{section:dataset} are partitioned so that 60\% is used to train the \acp{NN}, 20\% is used for validation and hyper-parameter tuning, and the remaining 20\% is used for testing. The \acp{NN} are trained for 2000 epochs with a batch size of 800 using the ADAM optimizer \cite{Kingma2015AdamAM}. The $\gamma$ parameter in the utility function is chosen to be 0.3 empirically to balance the \ac{SNR} and \ac{IA} performance. Operators may tune $\gamma$ individually for each deployment scenario as a hyperparameter. The measurement noise is \ac{AWGN}. For \ac{CB} baselines, the \ac{BS} and the \ac{UE} adopt over-sampled \ac{DFT} codebooks with 256 and 64 beams respectively. The rest of the simulation parameters are summarized in Table \ref{table:data_params}. The code for this work is published.\footnote{\url{https://github.com/YuqiangHeng/DLGF}}

\subsection{Gain of DL-GF vs. Baselines}
\subsubsection{Can DL-GF Achieve High SNR?}

\begin{figure}[!thp]
  \centering
  \includegraphics[width=0.775\textwidth]{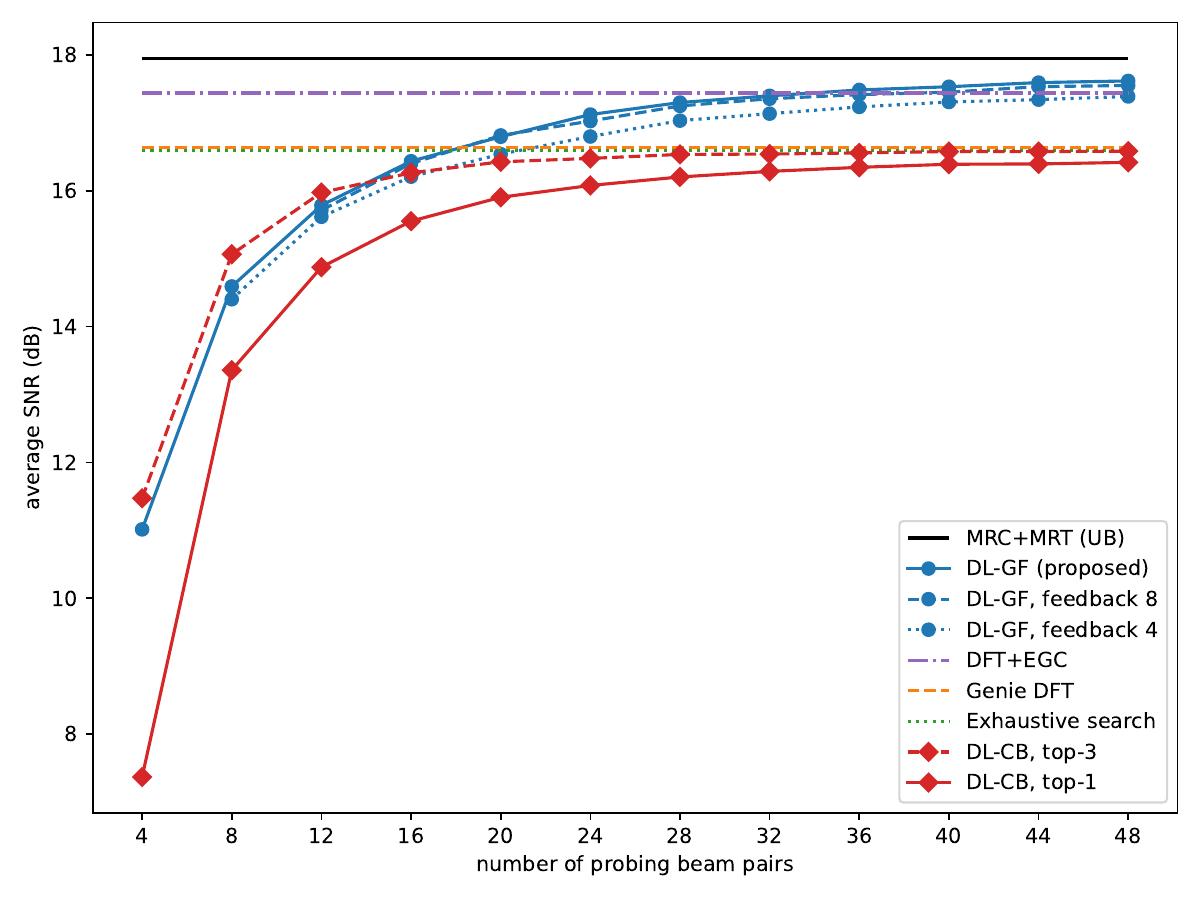}
  \caption{The average \ac{SNR} vs. number of probing beam pairs in the O1 scenario.}\label{figure:snr_vs_npb_O1}
\end{figure}

\begin{figure}[!thp]
  \centering
  \includegraphics[width=0.775\textwidth]{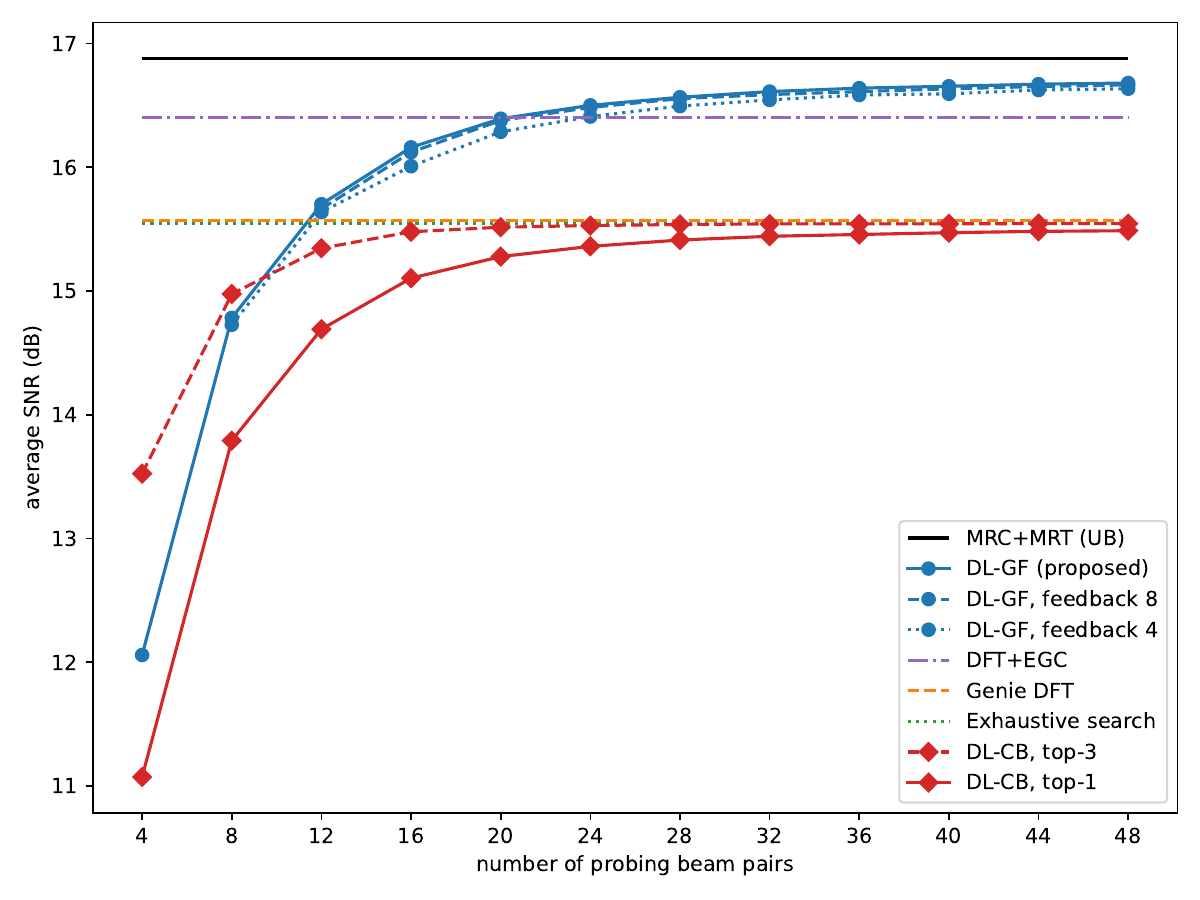}
  \caption{The average \ac{SNR} vs. number of probing beam pairs in the I3 BS 1 scenario.}\label{figure:snr_vs_npb_I3_BS1}
\end{figure}

\begin{figure}[!thp]
  \centering
  \includegraphics[width=0.775\textwidth]{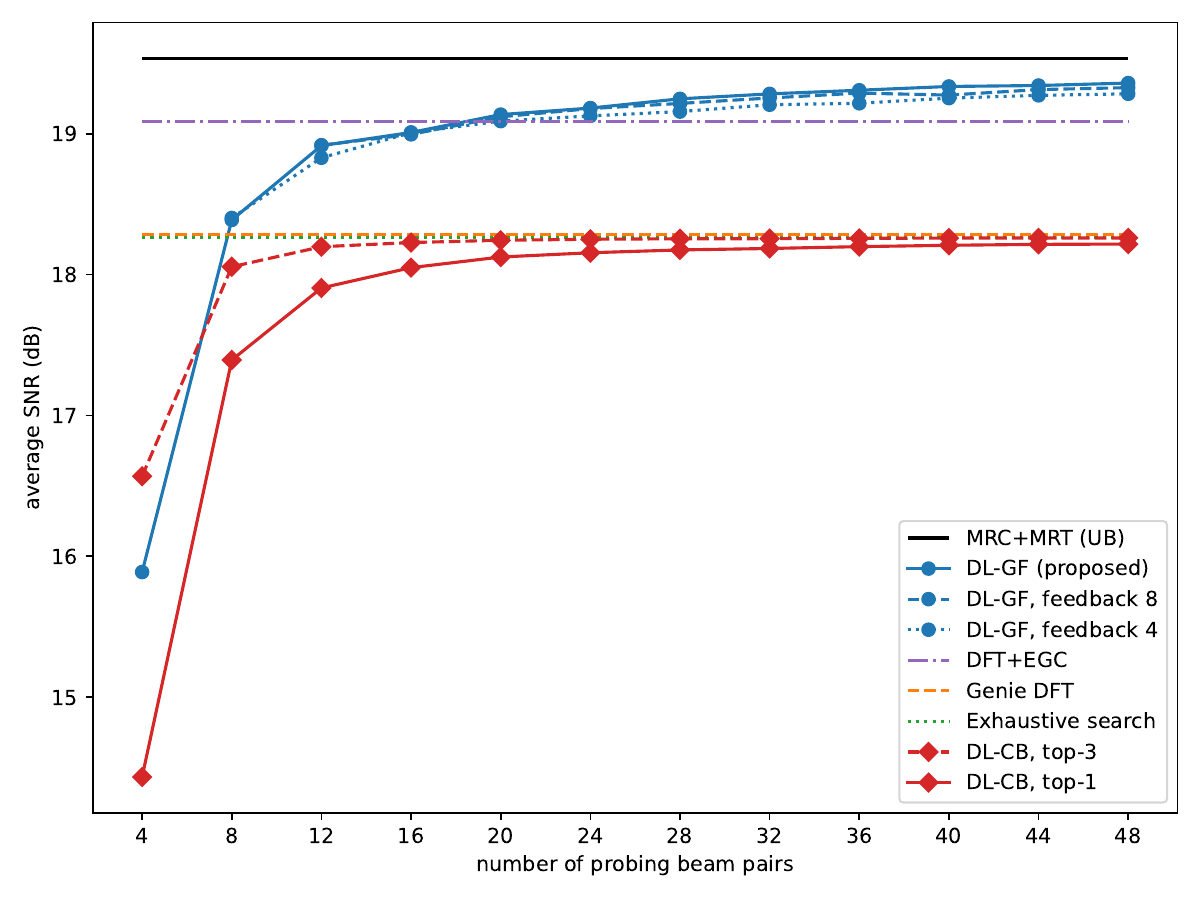}
  \caption{The average \ac{SNR} vs. number of probing beam pairs in the I3 BS 2 scenario.}\label{figure:snr_vs_npb_I3_BS2}
\end{figure}

An ideal beam alignment method should quickly select near-optimal beams and achieve high \ac{SNR}. The average \ac{SNR} achieved by the proposed DL-GF method and the baselines in the more realistic O1 and I3 scenarios are shown in Fig. \ref{figure:snr_vs_npb_O1}, \ref{figure:snr_vs_npb_I3_BS1}, \ref{figure:snr_vs_npb_I3_BS2}. It is evident that with the DL-CB approach, the beam classifiers can accurately select the best beam pairs. With an increasing number of probing beam pairs, the average \ac{SNR} achieved by the DL-CB baseline approaches that by the genie DFT. Nevertheless, its performance is limited by the resolution of the \ac{BS} and \ac{UE} codebooks since there is still a 1 dB gap between the DFT+EGC baseline and the genie DFT. On the other hand, the proposed DL-GF method improves upon DL-CB with its fully synthesized beams and is able to beat the exhaustive search with 20 probing beam pairs in the O1 scenario, 12 in I3 with BS 1 activated, and 8 in I3 with BS 2 activated. It is eventually able to outperform the DFT+EGC baseline with as few as 20 probing beam pairs in the I3 scenarios and 32 in O1. A comparison of the 10th, 50th, 90th percentile and the average \ac{SNR} with 32 probing beam pairs is shown in Table \ref{table:snr_distribution}. The proposed DL-GF method outperforms the exhaustive search everywhere in the distribution in both the O1 and the I3 scenario with \ac{BS} 1 activated, and is only slightly worse in the 10th percentile \ac{SNR} in the I3 scenario with \ac{BS} 2 activated. Meanwhile, the misdetection probability is just 0.117$\%$, 0.013$\%$ and 2.422$\%$ in the O1, I3 BS 1 and I3 BS 2 environments, guaranteeing that the vast majority of \acp{UE} can complete \ac{IA} with one of the probing beams. 

The gain of \ac{DL}-\ac{GF} is twofold. First, in the probing phase, the site-specific probing and sensing beams allow the \ac{BS} and the \ac{UE} to capture crucial channel information with fewer measurements. Second, in the beam prediction phase, the beam synthesizer functions can be intuitively viewed as infinitely large codebooks, whose resolution in practice is only limited by the floating-point precision of the \ac{NN} and the range of probing measurements. By generating analog beams at increased spatial resolutions that are also adapted to the specific environment, \ac{DL}-\ac{GF} achieves additional gain over the standard \ac{DFT} codebooks.
The proposed method can also be viewed to perform implicit channel estimation and beam alignment jointly. The probing phase is analogous to \ac{CS}, where a few learned site-specific beams are used to sense the channel. As a result, the proposed method can approach and even beat baselines such as DFT+EGC which require full \ac{CSI}.

\begin{table*}
\small
\centering
  \caption{SNR Comparison with 32 Probing Beam Pairs}\label{table:snr_distribution}
    \begin{tabular}{| c | c | c | c | c | c | c | c |}
    \hline
    \textbf{Scenario} & \textbf{SNR (dB)} & \textbf{DL-GF} & \makecell{\textbf{DL-CB}\\\textbf{top-3}} & \makecell{\textbf{Exhaustive}\\ \textbf{search}} & \textbf{Genie DFT} & \textbf{DFT+EGC} & \textbf{MRT+MRC}\\ \hline
\multirow{4}{*}{\makecell{O1\\(misdetection\\probability\\= 0.117\%)}} & 10th \% & 11.249 & 10.591 & 10.630 & 10.835 & 11.744 & 12.421 \\ \cline{2-8}
& 50th \% & 16.286 & 15.310 & 15.315 & 15.365 & 16.217 & 16.707 \\ \cline{2-8}
& 90th \% & 20.513 & 19.661 & 19.713 & 19.723 & 20.517 & 20.918 \\ \cline{2-8}
& average & 17.398 & 16.540 & 16.597 & 16.634 & 17.440 & 17.949 \\ \hline

\multirow{4}{*}{\makecell{I3, BS 1\\(misdetection\\probability\\= 0.013\%)}} & 10th \% & 13.804 & 12.535 & 12.535 & 12.624 & 13.805 & 14.316 \\ \cline{2-8}
& 50th \% & 16.513 & 15.351 & 15.353 & 15.385 & 16.269 & 16.771 \\ \cline{2-8}
& 90th \% & 18.349 & 17.472 & 17.472 & 17.480 & 18.170 & 18.518 \\ \cline{2-8}
& average & 16.611 & 15.541 & 15.542 & 15.572 & 16.402 & 16.879 \\ \hline

\multirow{4}{*}{\makecell{I3, BS 2\\(misdetection\\probability\\= 2.422\%)}} & 10th \% & 5.327 & 5.492 & 5.793 & 6.185 & 7.246 & 8.090 \\ \cline{2-8}
& 50th \% & 17.641 & 16.571 & 16.614 & 16.635 & 17.672 & 18.167 \\ \cline{2-8}
& 90th \% & 23.385 & 22.348 & 22.351 & 22.358 & 23.131 & 23.529 \\ \cline{2-8}
& average & 19.283 & 18.255 & 18.262 & 18.283 & 19.089 & 19.535 \\ \hline
    \end{tabular}
\end{table*}

\subsubsection{The \ac{SNR} vs. Beam Alignment Speed Trade-off}

\begin{figure}
  \centering
  \includegraphics[width=0.775\textwidth]{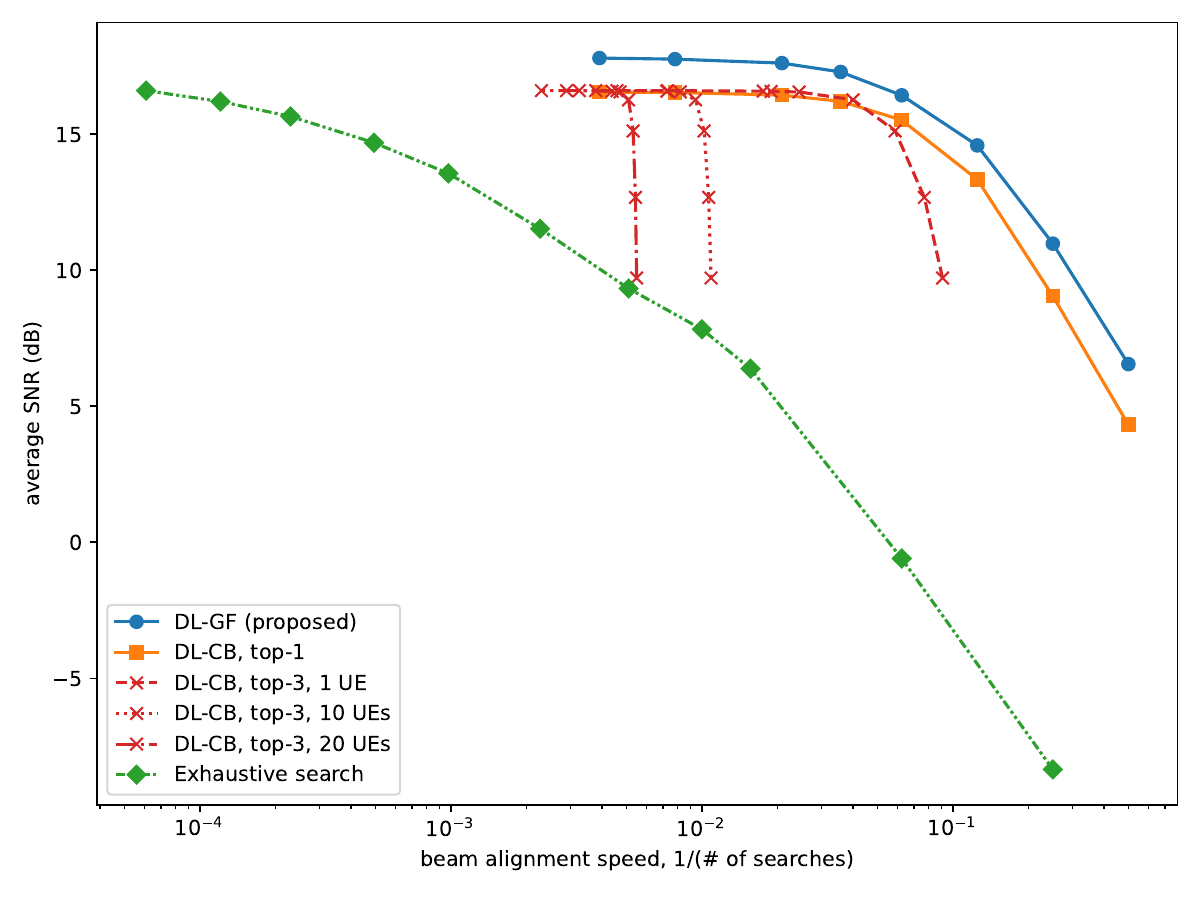}
  \caption{Average \ac{SNR} vs. the total beam alignment speed in the O1 scenario.}\label{figure:avg_snr_vs_ba_speed}
\end{figure}

Conventional \ac{CB} beam alignment approaches can always find better beams more frequently by trying more candidates and increasing the resolution of the codebook. Similarly, the proposed \ac{DL}-\ac{GF} method can achieve better \ac{SNR} by sweeping more probing beams. However, such \ac{SNR} gain usually occurs at the cost of higher latency. In practical cellular systems with multiple \acp{UE}, the beam alignment procedure may need to be performed for each \ac{UE}. It is therefore important for operators to consider the trade-off between the \ac{SNR} and the overall beam alignment speed, which we define as the reciprocal of the total number of beams swept for all \acp{UE}. 
The average \ac{SNR} achieved at each beam alignment speed for \ac{DL}-{GF}, \ac{DL}-\ac{CB} and the exhaustive search is shown in Fig. \ref{figure:avg_snr_vs_ba_speed}. Compared to the exhaustive search, the proposed \ac{DL}-\ac{GF} method is better by 5 to 10 dB in terms of the average \ac{SNR} at a given speed or faster by around two orders of magnitude when the achieved average \ac{SNR} is from 10 to 15 dB. The proposed \ac{GF} approach is also strictly better than the \ac{CB} method, achieving 1-2.5 dB higher \ac{SNR} at each beam alignment speed. With multiple \acp{UE}, other beam alignment methods that require an additional beam sweeping phase for each \ac{UE} are expected to achieve a similar \ac{SNR}-speed trade-off as the \ac{DL}-\ac{CB} with top-$k$ search does. Since \ac{DL}-\ac{GF} is faster than \ac{DL}-\ac{CB} with top-3 search by an order of magnitude with 10 or 20 \acp{UE}, similar gains can be expected over methods such as the hierarchical search and active learning approaches. Interestingly, increasing the number of probing beams is often more efficient than searching more candidates for the \ac{DL}-\ac{CB} method, particularly when a high beam alignment speed is required.

A more detailed analysis of the total number of beams swept for different beam alignment methods in a cell with $K$ \acp{UE} is shown in Table \ref{table:overhead}. In the hierarchical search, the \ac{BS} and the \ac{UE} adopt 2-tier codebooks with $N_{\mathbf{F}}$ and $N_{\mathbf{W}}$ wide beams respectively. The \ac{BS} first performs the \ac{Tx} hierarchical search while the \ac{UE} uses an omnidirectional \ac{Rx} beam, then the \ac{UE} performs the \ac{Rx} hierarchical search. The exhaustive search requires sweeping all combinations of the $M_{\textnormal{T}}$ Tx beams and $M_{\textnormal{R}}$ Rx beams, which amounts to a total of 16,384 beam pairs in our setting. Methods such as the hierarchical search and \ac{DL}-\ac{CB} with top-$k$ search require an additional beam sweeping phase for each \ac{UE}. As a result, the overall beam sweeping overhead increases linearly with the number of \acp{UE}. On the other hand, since the proposed \ac{DL}-\ac{GF} method predicts the beams in one shot after sweeping a common set of probing beam pairs for all \acp{UE}, its beam sweeping overhead is constant regardless of the number of \acp{UE}. In our setting, \ac{DL}-\ac{GF} reduces the beam sweeping overhead by over 500$\times$ while scaling optimally with increasing number of \acp{UE}.

\begin{table*}
\small
\centering
  \caption{Beam Sweeping Overhead for $K$ UEs}\label{table:overhead}
    \begin{tabular}{| c | c | c |}
    \hline
    \textbf{Beam alignment method} & \textbf{Beam sweeping overhead} & \textbf{Feedback overhead} \\ \hline
    DL-GF & $N_{\textnormal{probe}} $ &  $K N_{\textnormal{probe}}$ power values \\ \hline
    DL-CB, top-$k$ & $N_{\textnormal{probe}} + \min(K \cdot k_{\mathbbm{1}_{\{k>1\}}},M_{\textnormal{T}}M_{\textnormal{R}})$ & \makecell{$K N_{\textnormal{probe}}$ power values\\ + $K \cdot \mathbbm{1}_{\{k>1\}}$ beam indices} \\ \hline
    \makecell{Hierarchical search with\\$N_{\mathbf{F}},N_{\mathbf{W}}$ Tx and Rx wide beams} & $N_{\mathbf{F}} + \min(K \frac{M_{\textnormal{T}}}{N_{\mathbf{F}}},M_{\textnormal{T}}) + N_{\mathbf{W}} + \min(K \frac{M_{\textnormal{R}}}{N_{\mathbf{W}}},M_{\textnormal{R}}) $ & $4 K$ beam indices \\ \hline
    \makecell{Exhaustive search of\\$M_{\textnormal{T}},M_{\textnormal{R}}$ Tx and Rx beams}& $M_{\textnormal{T}}M_{\textnormal{R}}$ & $K$ beam indices\\ \hline
    \end{tabular}
\end{table*}

\subsection{Intuition Behind the Learned Beams}\label{section:probing_codebook_pattern}

\begin{figure*}[!htb]
    \centering
    \begin{subfigure}[b]{0.49\textwidth}
        \centering
      \begin{overpic}[width=\textwidth]{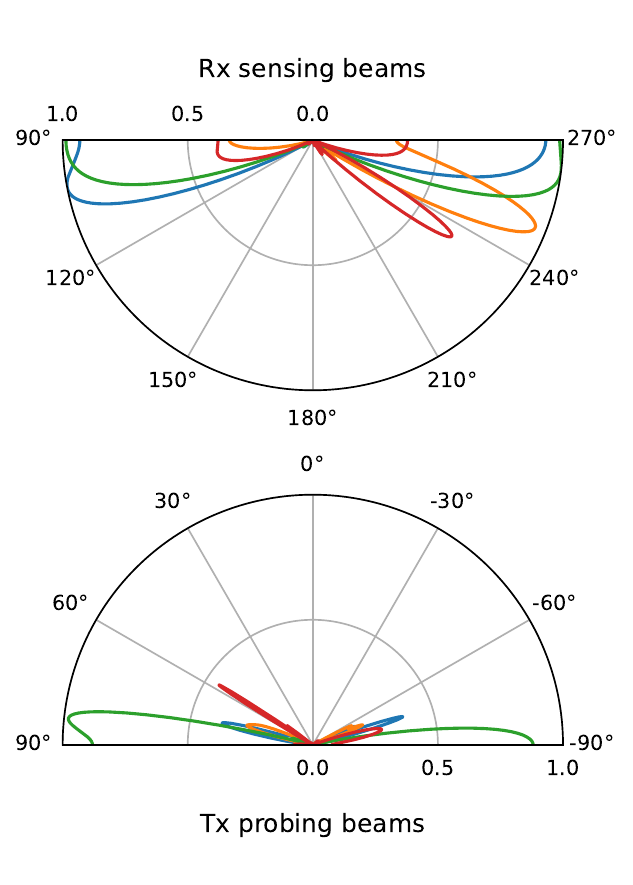}
        \put(0,160){\small \makecell{highly site-specific \\probing spatial coverage\\ to avoid blockage}}
      \end{overpic}
      \caption{Learned probing beam pairs}
      \label{figure:4_pb_probing}
    \end{subfigure}
    \hfill
    \begin{subfigure}[b]{0.49\textwidth}
    \centering
      \begin{overpic}[width=\textwidth]{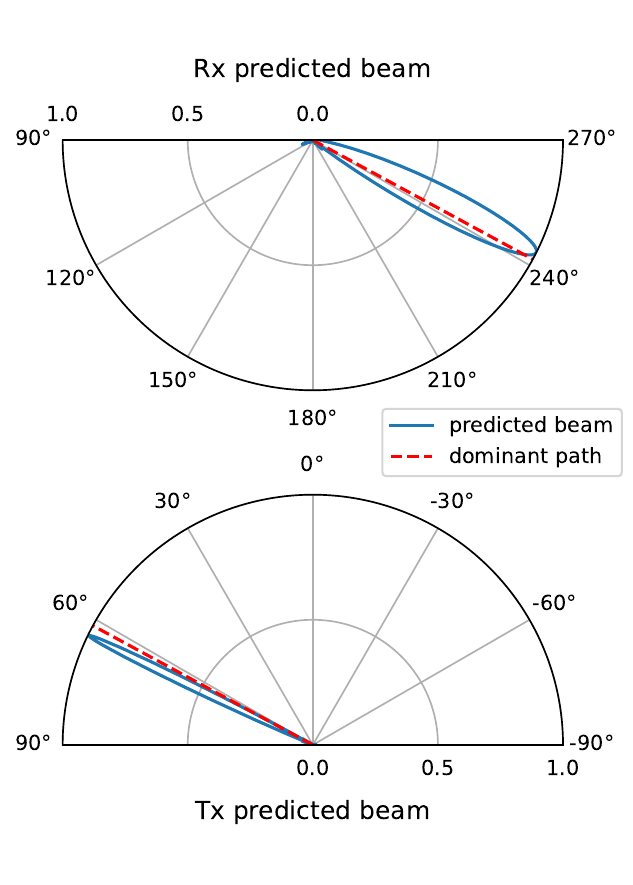}
        \put(0,160){\small \makecell{narrow predicted beams\\ slightly misaligned with\\ the dominant path}}
      \end{overpic}
      \caption{Predicted beam pair for communication}
      \label{figure:4_pb_predicted}
    \end{subfigure}
\caption{Radiation patterns of learned probing beam pairs and predicted beams in the O1 blockage scenario, $N_{\mathrm{probe}}=4$.}
\label{figure:codebook_pattern_4_probing_beam}
\end{figure*}

\begin{figure*}[!htb]
    \centering
    \begin{subfigure}[b]{0.49\textwidth}
        \centering
      \begin{overpic}[width=\textwidth]{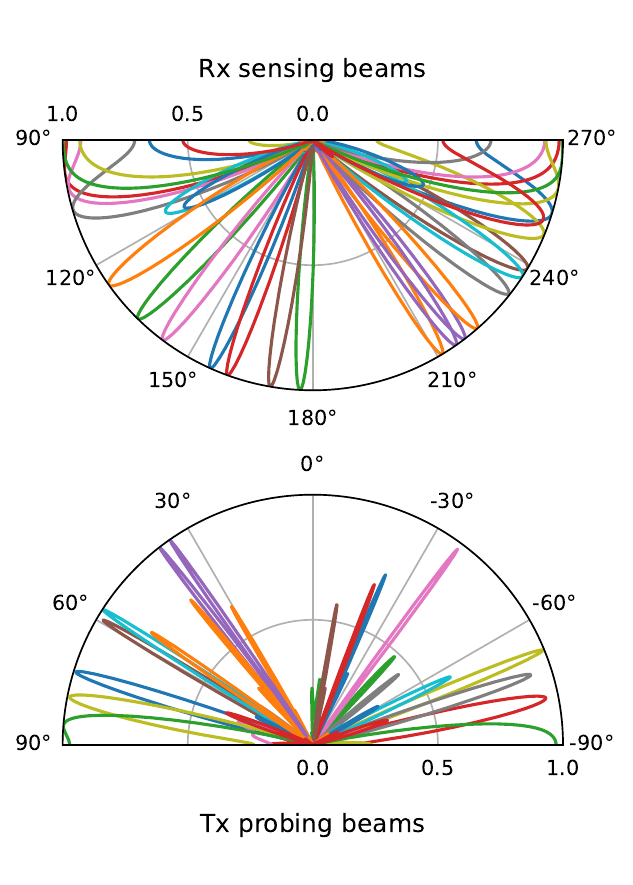}
        \put(20,160){\small \makecell{increased probing \\spatial coverage}}
      \end{overpic}
      \caption{Learned probing beam pairs}
      \label{figure:20_pb_probing}
    \end{subfigure}
    \hfill
    \begin{subfigure}[b]{0.49\textwidth}
    \centering
      \begin{overpic}[width=\textwidth]{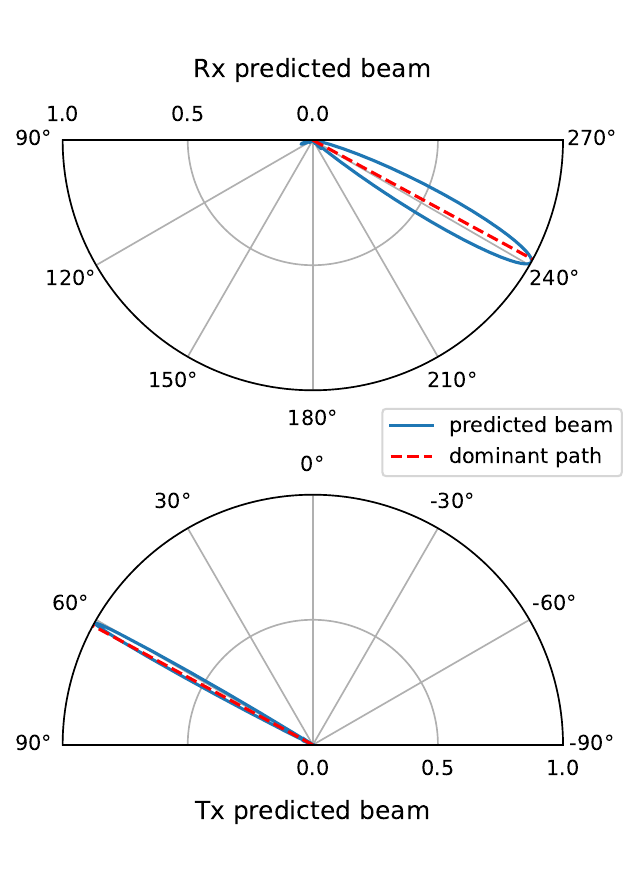}
        \put(0,160){\small \makecell{narrow predicted beams \\ closely aligned with\\ the dominant path}}
      \end{overpic}
      \caption{Predicted beam pair for communication}
      \label{figure:20_pb_predicted}
    \end{subfigure}
\caption{Radiation patterns of learned probing beam pairs and predicted beams in the O1 blockage scenario, $N_{\mathrm{probe}}=24$.}
\label{figure:codebook_pattern_20_probing_beam}
\end{figure*}

To better understand how the probing beam pairs and the beam synthesizer functions achieve better performance, we investigate the learned beam patterns in the O1 Blockage scenario without random \ac{UE} orientations. For easier visualization, the \ac{BS} and the \acp{UE} are equipped with \ac{ULA} arrays that only beamform in the azimuth domain. A \ac{UE} located on the left of the \ac{BS} is selected from the testing dataset as a case study.
With 4 probing beam pairs, the probing and sensing beam patterns are heavily adapted to the propagation environment by directing most of the energy towards the reflectors on both ends of the street while avoiding the blockage in the broadside direction, as shown in Fig. \ref{figure:codebook_pattern_4_probing_beam}. The probing and sensing beams often have multiple lobes, which likely allows them to capture channel information more effectively. 
On the other hand, the synthesizers predict sub-optimal beams. Both the \ac{Tx} and \ac{Rx} beams have a single narrow main lobe, but are misaligned with the dominant path by 2.51 and 1.79 degrees respectively.
As the number of probing beam pairs is increased to 24, the learned probing beams have much larger spatial coverage, as shown in Fig. \ref{figure:codebook_pattern_20_probing_beam}. This is particularly noticeable in the \ac{Rx} sensing beams since the range of \ac{AoA} is much larger compared to that of the \ac{AoD} with more variations in the location of \acp{UE}. Utilizing the increased information gathered by the probing beam pairs, the beam synthesizers also learn to focus energy more accurately. Both the predicted \ac{Tx} and \ac{Rx} beams are better aligned with the dominant path and are off by just 0.73 degrees. Overall, this allows \ac{DL}-\ac{GF} to achieve better \ac{SNR} by adopting more probing beams.

\subsection{Validating the Proposed Design}
\subsubsection{Benefits of Site-specific Probing} \label{section:ablation_probing_strategy}
\begin{figure}[!tbp]
  \centering
  \includegraphics[width=0.775\textwidth]{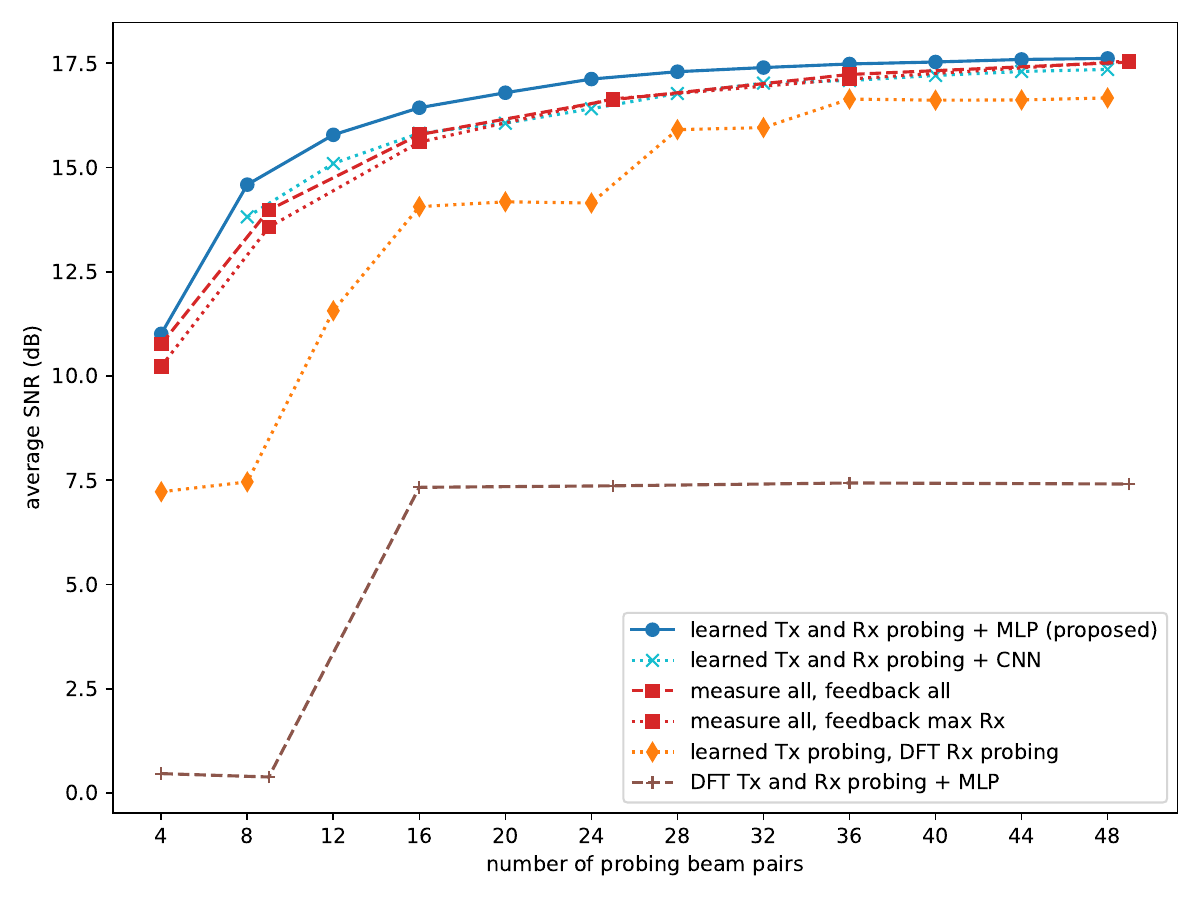}
  \caption{Average \ac{SNR} vs. the number of probing beam pairs with different probing beam designs, probing strategies and NN architectures in the O1 scenario.}\label{figure:avg_snr_vs_npb_ablation}
\end{figure}

The learned probing beam pairs provide important information about the channel for the downstream \ac{NN} beam synthesizers. To verify the benefits of site-specific probing, the learned probing and sensing beams are replaced with evenly spaced pencil beams from sub-sampled \ac{DFT} codebooks. Compared to learning site-specific probing beam pairs, the average \ac{SNR} drops by over 10 dB when both the \ac{BS} and the \ac{UE} adopt \ac{DFT} beams as shown in Fig. \ref{figure:avg_snr_vs_npb_ablation}. Note that in this setting, the \ac{UE} measures and reports all combinations of the probing and sensing beams so that the total number of probing beam pairs is $N_{\textnormal{probe}}^2$. If the \ac{BS} learns site-specific probing beams while the \ac{UE} adopts \ac{DFT} sensing beams, there is still a 2 dB loss in the average \ac{SNR}. Therefore, learning site-specific probing and sensing beams for the \ac{BS} and the \ac{UE} can provide significant gain. 

The proposed method learns a \ac{UE} beam for each \ac{BS} probing beam. An alternative strategy is to measure all combinations of probing and sensing beams and feed back all the measurements, which requires a total number of $N_{\textnormal{probe}}^2$ probing beam pairs instead of $N_{\textnormal{probe}}$ in the proposed design. As shown in Fig. \ref{figure:avg_snr_vs_npb_ablation}, this approach is consistently worse than the proposed design, particularly when the number of probing beam pairs is small. Feeding back only the maximum \ac{Rx} measurement for each \ac{BS} probing beam results in little further performance degradation when using more than 25 probing beam pairs. This indicates that most of the sensing measurements for each probing beam provide little additional information about the channel. Overall, the proposed probing strategy can capture information about the channel more efficiently with a limited number of probing measurements. 

\subsubsection{Verifying the Beam Synthesizer Architecture} \label{section:ablation_NN}
The \ac{NN} beam synthesizers use a fully connected \ac{MLP} architecture due to the relatively small feature dimension. For comparison, an alternative beam synthesizer is designed with 3 one-dimensional convolutional layers followed by max pooling with a stride of 2 and a kernel size of 2, flattening, and a final fully connected output layer. Each convolutional layer has 64 channels, a kernel size of 3 and \ac{ReLU} activation. As shown in Fig. \ref{figure:avg_snr_vs_npb_ablation}, the proposed design consistently achieves better average \ac{SNR} than the \ac{CNN}-based design does, indicating that the \ac{MLP} is an appropriate architecture in this setting.  

\subsubsection{Reducing the Misdetection Probability}\label{section:misdetection_prob}
\begin{figure}[!tbp]
  \centering
  \includegraphics[width=0.775\textwidth]{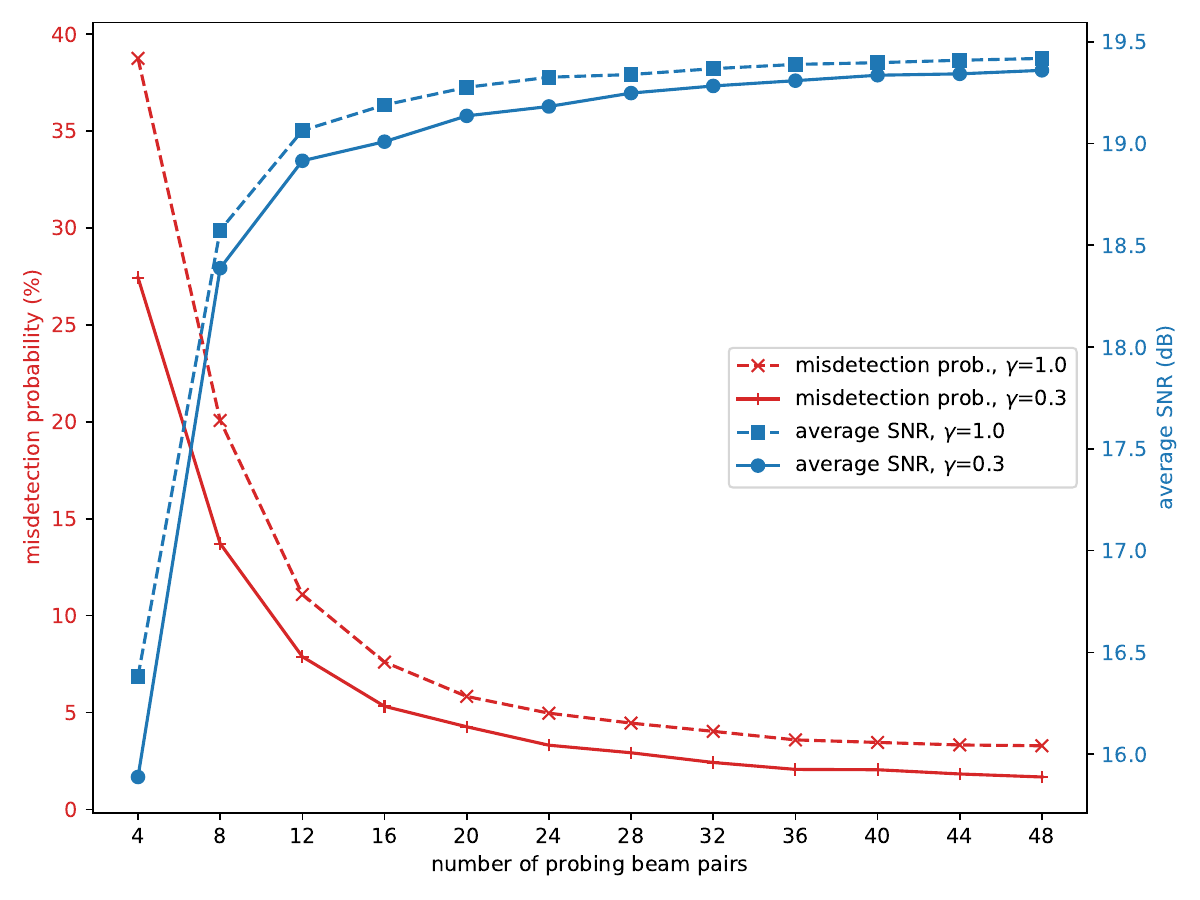}
  \caption{The misdetection probability and average \ac{SNR} vs. the number of probing beam pairs with different $\gamma$ during training in the I3 scenario with BS 2 activated.}\label{figure:misdetection_prob_vs_npb_vs_gamma}
\end{figure}

\begin{figure}[!tbp]
  \centering
  \includegraphics[width=0.775\textwidth]{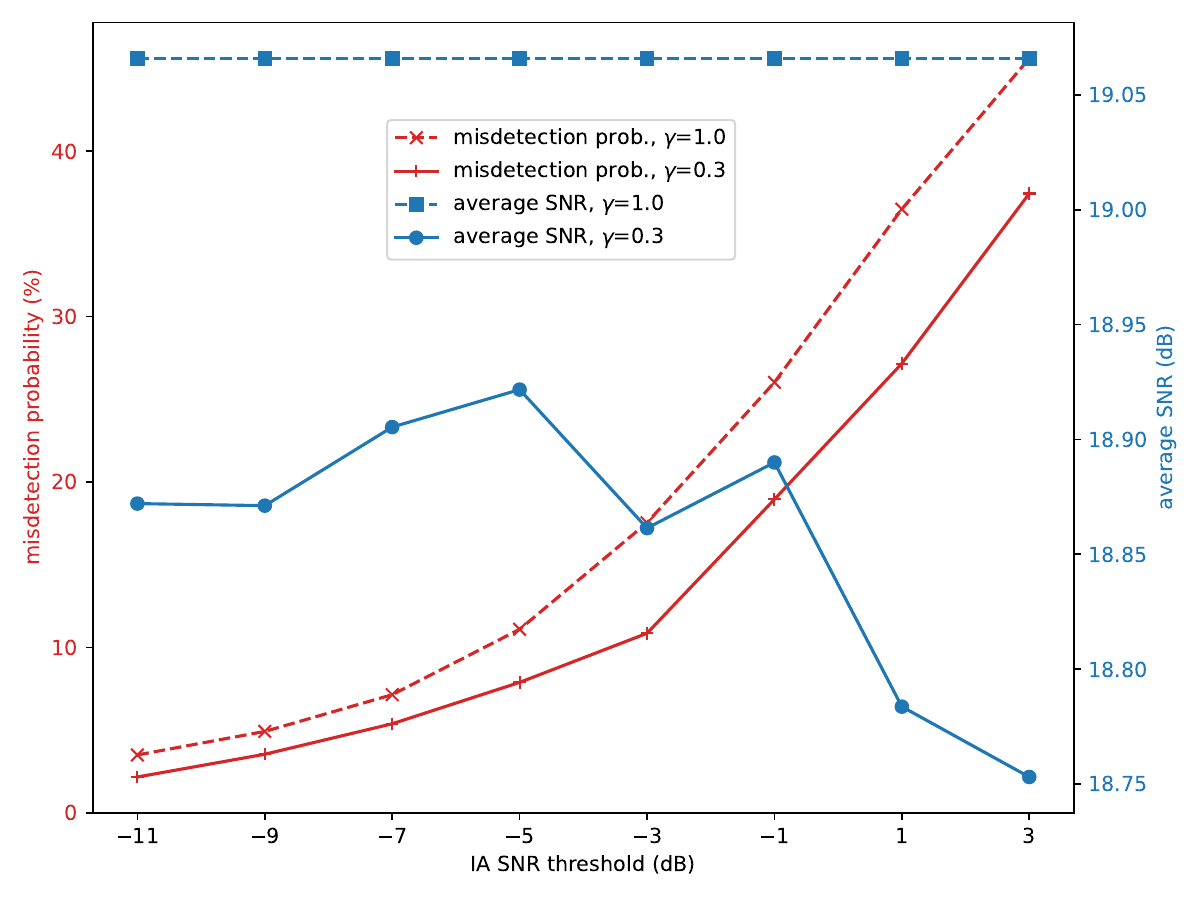}
  \caption{The misdetection probability and average \ac{SNR} vs. the \ac{IA} \ac{SNR} threshold with different $\gamma$ in the I3 scenario with BS 2 activated. The proposed method uses 12 probing beam pairs.}\label{figure:misdetection_prob_vs_IA_threshold_vs_gamma}
\end{figure}

The ideal beam alignment solution should not only achieve high \ac{BF} gain for connected \acp{UE}, but also be able to discover new \acp{UE} and allow them to complete \ac{IA}. With the proposed DL-GF method, a new \ac{UE} can do so with one of the probing beam pairs if it satisfies a minimum \ac{SNR} requirement. Naturally, adopting more probing beams will allow each to cover a smaller angular space, thus increasing the gain of the probing beams and reducing the misdetection probability. By explicitly incorporating the \ac{IA} performance in the utility function, we provide another tuning nob to reduce the misdetection probability.  
The proposed utility function can indeed further reduce the misdetection probability when the number of probing beam pairs is fixed, as shown in Fig. \ref{figure:misdetection_prob_vs_npb_vs_gamma}. Compared to simply optimizing the \ac{BF} gain of the synthesized beams ($\gamma=1.0$), incorporating the \ac{IA} performance ($\gamma=0.3$) reduces the misdetection probability by 6.3 percentage points with 8 probing beam pairs and by 3.2 percentage points with 12 probing beam pairs while suffering from little \ac{SNR} loss. Although the gain diminishes with more probing beams, the two-component utility function still provides a powerful tool to improve the \ac{IA} coverage when the number of probing beams is limited. Since the $\mathcal{U}_{\textnormal{IA}}$ term only covers \acp{UE} below a specified \ac{SNR} threshold, the proposed utility function also allows the \ac{NN} model to adapt to different \ac{SNR} requirements for \ac{IA}. As shown in Fig. \ref{figure:misdetection_prob_vs_IA_threshold_vs_gamma}, the misdetection probability is reduced at various \ac{SNR} threshold values for \ac{IA} compared to simply optimizing the \ac{BF} gain. The gain is more significant if a higher \ac{SNR} threshold is required.

\subsection{Robustness of DL-GF}
\subsubsection{Robustness to Noisy Measurements}\label{section:measurement_noise}

\begin{figure}[!tbp]
  \centering
  \includegraphics[width=0.775\textwidth]{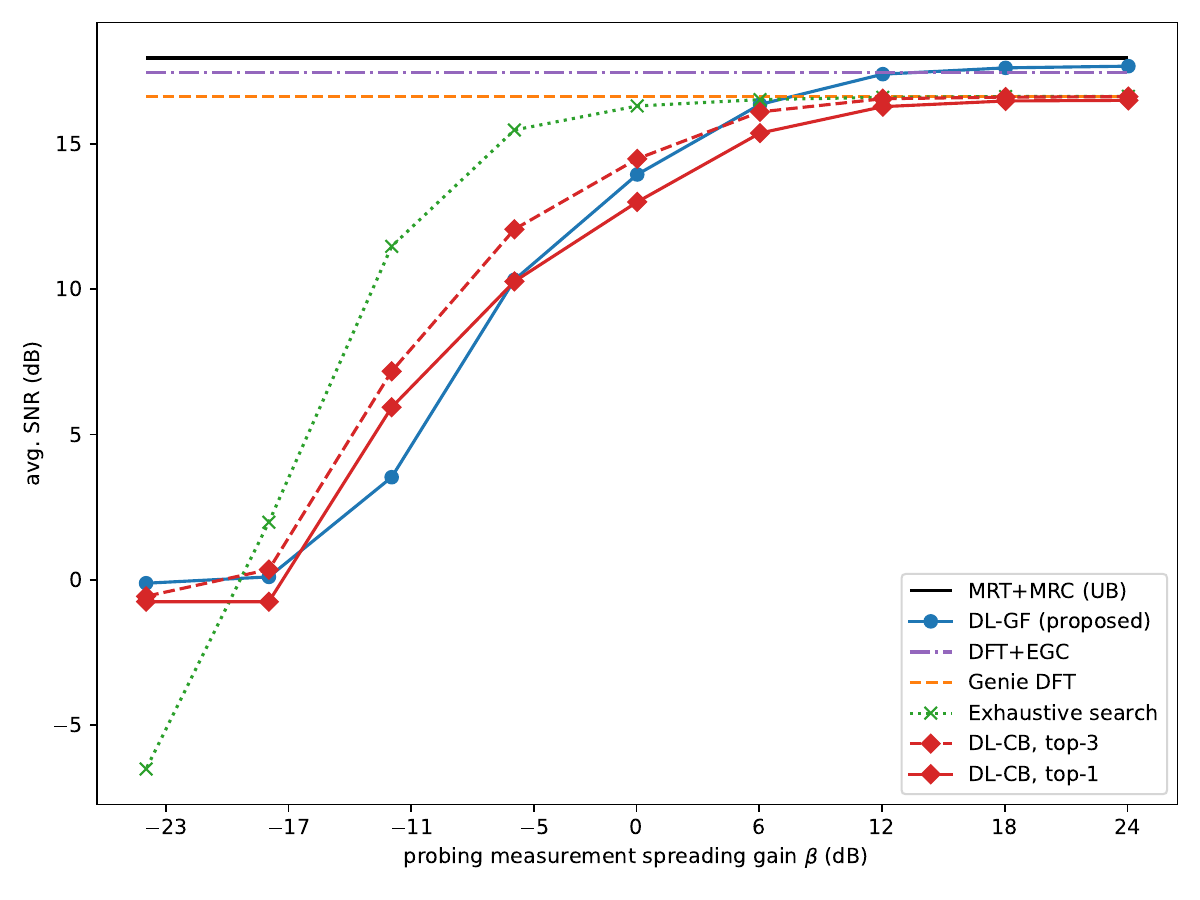}
  \caption{Comparison of the average \ac{SNR} with increasing spreading gain in the probing measurements in the O1 scenario. The DL-GF and DL-CB methods have 32 probing beam pairs. A negative spreading gain means measurements during probing are noisier than in data transmission. }\label{figure:avg_snr_vs_measurement_noise}
\end{figure}

The beam synthesizers rely on the probing measurements to generate the \ac{BF} weights. While measurement noise can cause search errors in exhaustive and hierarchical searches, it can also impact the prediction of the \ac{NN} models. A comparison of the average \ac{SNR} achieved by the predicted beams with increasing probing measurement spreading gain is shown in Fig. \ref{figure:avg_snr_vs_measurement_noise}. Both \ac{DL}-based methods are trained at each probing measurement spreading gain level. As expected, the performance of the DL-GF, DL-CB and exhaustive search improves with increasing measurement spreading gain. The exhaustive search benefits less: its performance plateaus with almost no measurement spreading gain. On the other hand, the proposed DL-GF and the DL-CB methods are less robust to measurement noise due to their probing beams having lower gain. They both require a spreading gain of around 12 dB in the probing measurements to achieve their best performance. To further investigate the impact of noise, negative spreading gain is considered, which implies a setting where the measurements are noisier during probing than in data transmission. Interestingly, the average \ac{SNR} of the exhaustive search drops sharply and becomes the worst when the measurements are extremely noisy. At this point, the exhaustive search essentially produces random guesses since the beam measurements are dominated by noise. On the other hand, the \ac{DL}-based methods converge to always predict the same beam pair. In the case of DL-CB, the predicted beam pair happens to be the most frequent optimal beam pair.

\subsubsection{Robustness to Imperfect Training Data}\label{section:imperfect_training_data}
\begin{figure}[!tbp]
  \centering
  \includegraphics[width=0.775\textwidth]{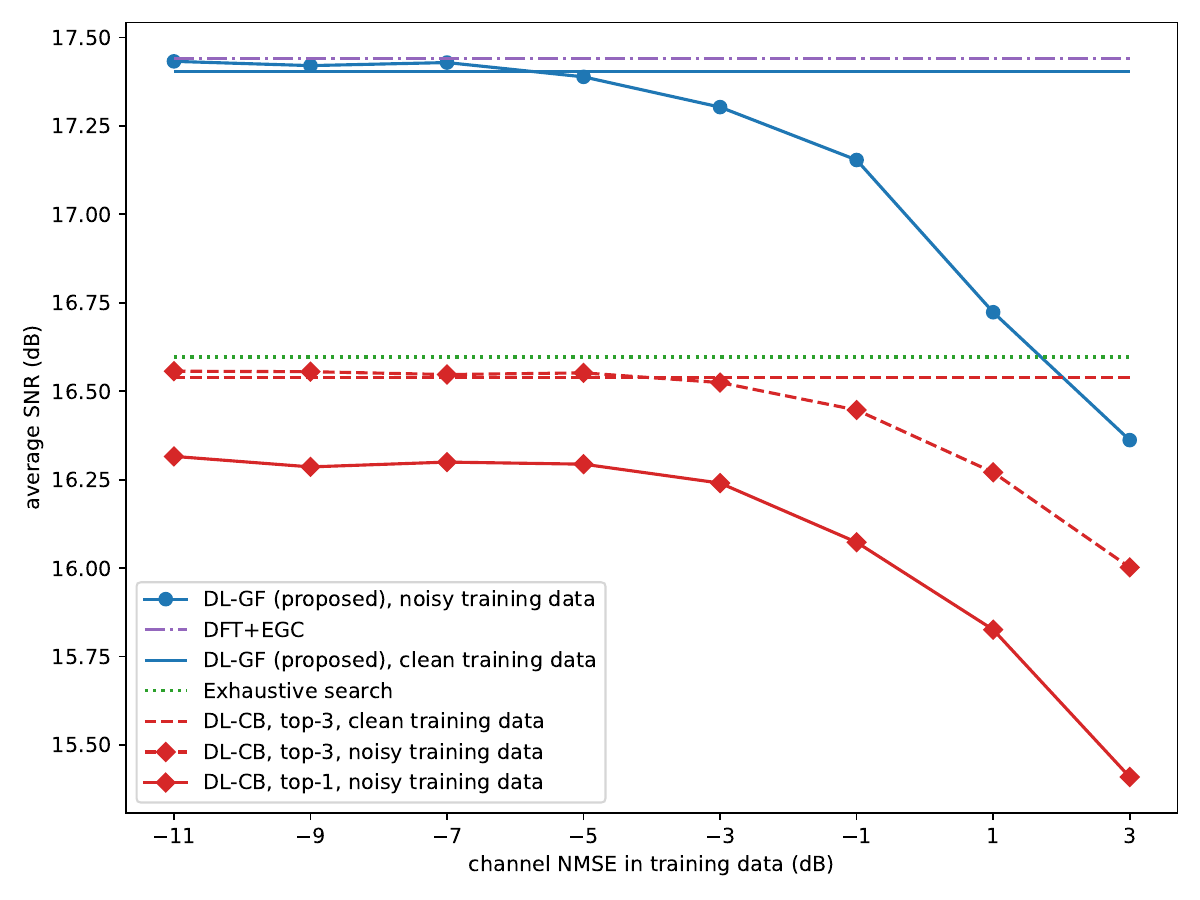}
  \caption{Comparison of the average \ac{SNR} with increasing training channel \ac{NMSE} in the O1 scenario. The DL-GF and DL-CB methods have 32 probing beam pairs.}\label{figure:avg_snr_vs_channel_NMSE}
\end{figure}

Training \ac{DL} models typically requires high-quality data. While ray-tracing is one of the most accurate ways to simulate \ac{mmWave} channels, it is not easily scalable to a city-wide deployment and the generated data may still differ from actual channels due to mismatched environments and imperfect propagation models. If the training data is acquired through measurement campaigns, noisy measurement and channel estimation errors would also affect the quality of the data. It is therefore important to investigate the performance of the proposed method when there is a mismatch between the channels used for training and the actual channels during deployment. To simulate imperfect training data, the training channel matrices are corrupted with \ac{AWGN} while the model is tested on clean channel data. The average \ac{SNR} of the proposed DL-GF method, the DL-CB baseline and the exhaustive search with increasing \ac{NMSE} in the training data is shown in Fig. \ref{figure:avg_snr_vs_channel_NMSE}. Both the DL-GF and the DL-CB methods are relatively robust to noisy training data, experiencing little performance degradation when the channel \ac{NMSE} is smaller than -3 dB. A small amount of noisy in the training data actually benefits the proposed DL-GF method, which is a common phenomenon in \ac{ML} where noisy training data can sometimes improve the robustness and generalizability of \ac{ML} models \cite{Goodfellow-et-al-2016}. Even with a channel \ac{NMSE} of 1 dB, the proposed DL-GF method can still outperform the exhaustive search and the DL-CB baseline trained on clean data. The added noise likely does not fundamentally shift the distribution of channels but instead makes it more ``fuzzy''. During the unsupervised training procedure, the proposed \acp{NN} still learn to beamform on these noisy channels and generalize well to the clean actual channels.

\subsubsection{Reducing the Feedback Overhead}
The gain of DL-GF can be partially attributed to the much richer uplink feedback: it uses all the measurements as feature vectors to synthesize the \ac{BF} weights instead of simply selecting the strongest beam. The feedback overhead of different methods are summarized in Table \ref{table:overhead}. To reduce the feedback overhead, \acp{UE} can report measurements of a few strongest probing beam pairs so that \acp{UE} use all the measurements to synthesize their beams while the \ac{BS} uses only the reported measurements.
The \ac{Rx} beam synthesizer $f_{\textnormal{R}}$ takes the full measurement vector $\mathbf{z}$ as input. The \ac{Tx} beam synthesizer $f_{\textnormal{T}}$ takes a masked version of $\mathbf{z}$ so that only the top-$m$ highest measurements are non-zero. Since the shape of the feature vector is roughly maintained, reducing the feedback only results in a small degradation in the average \ac{SNR}, as shown in Fig. \ref{figure:snr_vs_npb_O1}, \ref{figure:snr_vs_npb_I3_BS1}, \ref{figure:snr_vs_npb_I3_BS2}. For instance, in the O1 scenario with 32 probing beam pairs in total, the average \ac{SNR} drops by 0.263 dB when reporting 4 beam pairs and by only 0.043 dB when reporting the best 8. 

\subsubsection{Impact of Random UE Orientations}
\begin{figure}[!tbp]
  \centering
  \includegraphics[width=0.775\textwidth]{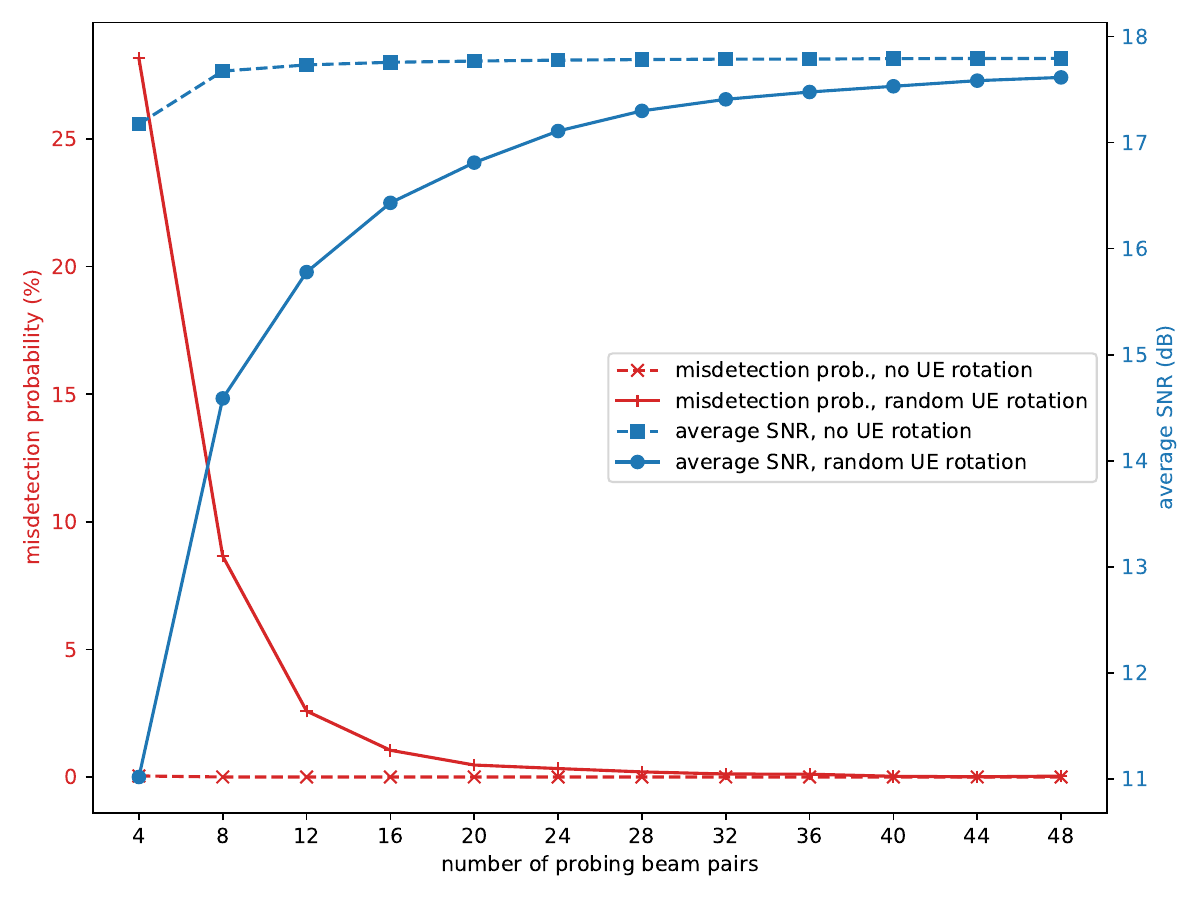}
  \caption{The average SNR and misdetection probability of the proposed DL-GF method with and without random UE orientations in the O1 scenario.}\label{figure:avg_snr_misdetection_prob_vs_random_rotation}
\end{figure}

The orientation of \acp{UE} has a large impact on the performance of the proposed DL-GF method. A comparison of the average \ac{SNR} and the misdetection probability with and without random \ac{UE} rotation in the O1 scenario is shown in Fig. \ref{figure:avg_snr_misdetection_prob_vs_random_rotation}. Without random \ac{UE} rotation, the proposed method can achieve better average \ac{SNR} and lower misdetection probability with significantly fewer probing beam pairs. The random \ac{UE} rotation increases the effective range of \ac{AoA}. As a result, more probing beam pairs are required to capture sufficient channel information on the \ac{UE} side. The \ac{UE}'s sensing beams also need to distribute energy and cover a larger angular space, leading to reduced \ac{BF} gain and worse misdetection probability.

\section{Conclusion}\label{section:conclusion}
The proposed \ac{GF} approach provides a promising new paradigm for \ac{mmWave} beam alignment by achieving better \ac{SNR} while being faster than existing methods by orders of magnitude. Its gain can be attributed to the combination of site-specific adaptation and the joint optimization of channel probing and beam synthesis.
Next-generation cellular systems can reap the benefits without overhauling the existing beam sweeping-based framework: the probing beams can be transmitted using periodic \acp{RS} while traditional beamforming codebooks are replaced with \ac{NN} beam synthesizers. Operators can select the number of probing beam pairs based on the \ac{SNR} and \ac{IA} coverage requirement for each site. 

There are many possible directions that this line of research could be extended, many of which have been identified throughout the paper. For example, scaling the site-specific training to city-wide networks and online adaptation to dynamic environments remain open research problems. 
Multi-beam and multi-\ac{UE} communication including the joint optimization of analog and digital \ac{BF} is another promising direction.
The beam alignment problem has mostly been investigated to date from the BS point of view. It is important to also consider the unique beam alignment challenges on the \ac{UE} side arising from multiple antenna panels, precarious dynamic rotations and more demanding power management.

\bibliographystyle{IEEEtran}
\bibliography{refs}

\end{document}